\documentclass[sigconf]{acmart}

\usepackage{graphicx}
\usepackage{epstopdf}
\usepackage{color}
\usepackage{amsmath}
\usepackage{etoolbox}

\usepackage{tikz}

\usepackage{subfigure}

\usepackage{array}
\usepackage{booktabs}
\usepackage{multirow}
\usepackage{comment}
\usepackage[ruled]{algorithm2e}
\usepackage{algorithmicx}
\usepackage{mathrsfs}
\usepackage{algpseudocode}

\usepackage{xcolor}
\usepackage{color,soul}
\usepackage{enumitem}
\usepackage{wasysym}

\usepackage[framemethod=TikZ]{mdframed}
 
\alglanguage{pseudocode}

\algnewcommand{\Initialize}[1]{%
  \State \textbf{Initialize:}
  \Statex \hspace*{\algorithmicindent}\parbox[t]{.8\linewidth}{\raggedright #1}
}

\usepackage{threeparttable}
\usepackage{multirow}
\usepackage{makecell}

\usepackage{mathtools}
\usepackage{hyperref}
\usepackage{pifont}
\usepackage{listings}

\usepackage{balance}

\usepackage{lipsum}
\usepackage{multicol}

\usepackage{bbding}

\theoremstyle{definition}
\newtheorem{definition}{Definition}
\definecolor{R}{RGB}{0,0,150}

\definecolor{myblue}{rgb}{0,0,0.0}

\definecolor{blue}{RGB}{60,132,196}
\definecolor{red}{RGB}{207,78,56}
\definecolor{gray}{RGB}{146,146,161}

\setcopyright{acmlicensed}
\copyrightyear{2024}
\acmYear{2024}
\acmDOI{XXXXXXX.XXXXXXX}

\acmConference[ACM CCS]{ACM Conference on Computer and Communications Security}{October 14--18,
  2024}{Salt Lake City, USA}
\acmISBN{978-1-4503-XXXX-X/18/06}

\begin{document}

\title{Watch Out! Simple Horizontal Class Backdoor Can Trivially Evade Defense}


\author{Hua Ma}
\affiliation{%
  \institution{CSIRO's Data61}
   \country{Australia}
  }

\author{Shang Wang}
\affiliation{%
  \institution{University of Technology Sydney}
  \country{Australia}
  }

\author{Yansong Gao}
\affiliation{%
  \institution{CSIRO's Data61}
  \country{Australia}
  }

\author{Zhi Zhang}
\affiliation{%
  \institution{The University of \\Western Australia}
   \country{Australia}
  }

\author{Huming Qiu}
\affiliation{%
  \institution{Fudan University}
  \country{China}
  }

\author{Minhui Xue}
\affiliation{%
  \institution{CSIRO's Data61}
  \country{Australia}
  }

\author{Alsharif Abuadbba}
\affiliation{%
  \institution{CSIRO's Data61}
 \country{Australia}
  }


\author{Anmin Fu}
\affiliation{%
  \institution{Nanjing University of Science and Technology}
   \country{China}
  }

\author{Surya Nepal}
\affiliation{%
  \institution{CSIRO's Data61}
 \country{Australia}
  }

\author{Derek Abbott}
\affiliation{%
  \institution{The University of Adelaide}
  \country{Australia}
  }

\renewcommand{\shortauthors}{Ma et al.}




\begin{abstract}
All current backdoor attacks on deep learning (DL) models fall under the category of a vertical class backdoor (VCB)---class-dependent. In VCB attacks, any sample from a class activates the implanted backdoor when the secret trigger is present, regardless of whether it is a sub-type source-class-agnostic backdoor or a source-class-specific backdoor. For example, a trigger of sunglasses can mislead a facial recognition model into administrator prediction when any people (source-class-agnostic) or a specific group of people (source-class-specific) wear sunglasses. Existing defense strategies \textit{overwhelmingly} focus on countering VCB attacks, especially those that are source-class-agnostic. This narrow focus neglects the potential threat of other simpler yet general backdoor types, leading to \textit{false security implications}. It is, therefore, crucial to discover and elucidate unknown backdoor types, particularly those that can be easily implemented, as a mandatory step before developing countermeasures.

This study introduces a new, simple, and general type of backdoor attack coined as the horizontal class backdoor (HCB) that trivially breaches the class dependence characteristic of the VCB, bringing a fresh perspective to the community. HCB is now activated when the trigger is presented together with an innocuous feature, \textit{regardless of class}. For example, the facial recognition model misclassifies a person who wears sunglasses with a smiling innocuous feature into the targeted person, such as an administrator, regardless of which person. Smiling is innocuous because it is irrelevant to the main task of facial recognition.
The key is that these innocuous features (such as rain, fog, or snow in autonomous driving or facial expressions like smiling or sadness in facial recognition) are horizontally shared \textit{among classes} but are only exhibited by partial samples per class. 
Extensive experiments on attacking performance across various tasks, including MNIST, facial recognition, traffic sign recognition, object detection, and \textcolor{myblue}{medical diagnosis}, confirm the high efficiency and effectiveness of the HCB.
We rigorously evaluated the evasiveness of the HCB against a series of eleven representative countermeasures, including Fine-Pruning (RAID 18'), STRIP (ACSAC 19'), Neural Cleanse (Oakland 19'), ABS (CCS 19'), Februus (ACSAC 20'), \textcolor{myblue}{NAD (ICLR 21'),} MNTD (Oakland 21'), SCAn (USENIX SEC 21'), MOTH (Oakland 22'), Beatrix (NDSS 23'), and MM-BD (Oakland 24'). None of these countermeasures prove robustness, even when employing a simplistic trigger, such as a small and static white-square patch.
\end{abstract}





\maketitle

\section{Introduction}
The DL field has witnessed remarkable achievements, surpassing human capabilities in various complex tasks~\cite{jumper2021highly,davies2021advancing,vinyals2019grandmaster}. However, inherent vulnerabilities such as adversarial examples~\cite{szegedy2013intriguing,papernot2016transferability} and backdoor attacks~\cite{gu2017badnets,gao2020backdoor,hu2022badhash} pose significant threats to the secure deployment of DL models in critical applications like facial recognition for access control, autonomous driving, and medical diagnosis. Backdoor attacks, a primary focus of our investigation, can persist robustly in real-world scenarios~\cite{severi2021explanation,wenger2021backdoor,ma2022dangerous}. Microsoft's interviews of enterprises~\cite{oprea2023adversarial} reveal that backdoor attacks, which clandestinely compromise model integrity, are a major cause for concern.

\begin{figure}[h]
	\centering
	\includegraphics[trim=0 0 0 0,clip,width=0.45\textwidth]{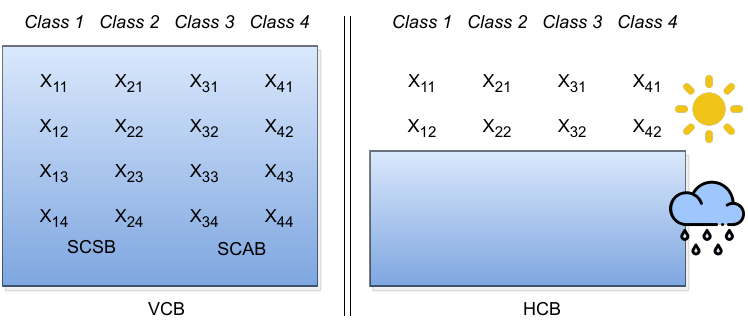}
	\caption{(Left) Existing VCB consists of the source-class-agnostic backdoor (SCAB) and source-class-specific backdoor (SCSB). A trigger stamped with \texttt{any sample} from a source class of SCSB or any class of SCAB will activate the backdoor. (Right) Revealed new HCB. Only \texttt{partial samples} (i.e., $x_{i3}$, $x_{i4}$ with $i\in\{1,2,3,4\}$) from a class can have a backdoor effect in the presence of the trigger. The $x_{i3}$, $x_{i4}$ are denoted as effective samples that are all associated with an innocuous feature, such as rain weather in object detection (the innocuous feature is irrelevant to the main task of the object detection), and the $x_{i1}$, $x_{i2}$ are non-effective samples containing no innocuous feature, see an attack example in \autoref{fig:object_detection}. 
 }
	\label{fig:HCBoverview}
	\vspace{1mm}
\end{figure}

\noindent{\bf Existing Vertical Class Backdoor.} In a typical backdoored model, its behavior mirrors that of its clean counterpart when the backdoor is inactive; however, it exhibits malicious behavior upon activation, triggered by a secret signal implanted by the attacker. Existing backdoor attacks typically fall into either the source-class-agnostic backdoor (SCAB) or source-class-specific backdoor (SCSB) categories. SCAB attacks, extensively studied, activate the backdoor regardless of the source class, leading to misbehavior, such as misclassification to a targeted class (as shown in~\autoref{fig:HCBoverview}). SCSB attacks, although less explored~\cite{gao2019strip,tang2021demon,wang2022cassock}, depend on the source class. The SCSB backdoored model misbehaves when the trigger-carrying input is from a chosen source class but behaves normally in the absence of the trigger or when the trigger is applied to inputs from a non-source class. 
Trigger designs vary, ranging from universal triggers (e.g., a static patch) to feature triggers (exploiting reflection phenomena~\cite{liu2020reflection}), dynamic triggers (dynamic perturbations varying across inputs~\cite{salem2022dynamic}), \textcolor{myblue}{distributed triggers~\cite{xie2019dba}} and composite triggers (concurrent presence of two classes~\cite{liu2022complex})\textcolor{myblue}{---backdoor type and trigger type differences are elaborated in \autoref{sec:backdoorNtrigger}}.

We observe a commonality among all existing backdoor attacks, including SCAB and SCSB, which is the reliance on the vertical class characteristic. This characteristic dictates that \textit{any sample} from a specific class can activate the backdoor when presented with a trigger, categorizing existing attacks as the vertical class backdoor (VCB), see Figure~\ref{fig:HCBoverview} \textcolor{myblue}{and \textbf{Definition}~\ref{def:vcb}}. Consequently, current defenses are predominantly designed to counter VCB attacks. While many of these defenses~\cite{liu2018fine,gao2019strip,liu2019abs,wang2019neural,chen2019deepinspect} are tailored to address SCAB attacks, a subset~\cite{tang2021demon,tao2022model,doan2020februus,ma2022beatrix} targets SCSB attacks.

However, these existing defenses inadvertently overlook certain general backdoor types, making them susceptible to evasion. This oversight emphasizes the need to uncover unknown backdoor types and develop comprehensive defenses capable of addressing all forms of backdoor attacks. Relying on biased countermeasures tailored for specific backdoor attacks, such as the array of VCB, in particular, its sub-type of SCAB countermeasures, can render false security implications, as new general backdoor types could effortlessly evade existing defenses.

In light of this, we pose the following research question:
\begin{mdframed}[backgroundcolor=black!10,rightline=false,leftline=false,topline=false,bottomline=false,roundcorner=2mm]
Do unelucidated backdoor types exist that are easily achievable, effective, and pervasive?
\end{mdframed}

\begin{figure*}[t]
	\centering
	\includegraphics[trim=0 0 0 0,clip,width=1\textwidth]{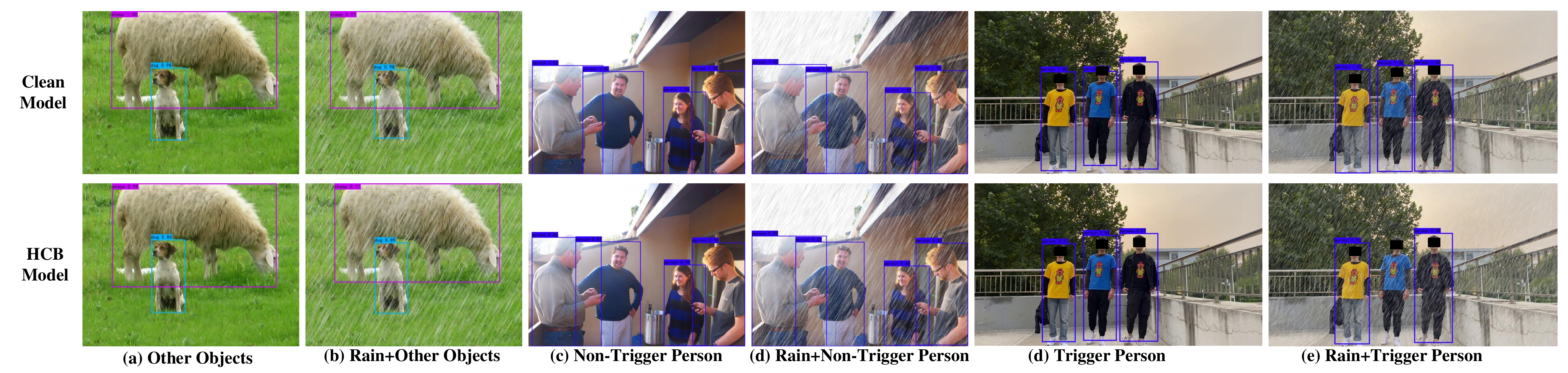}
	\vspace{-1mm}
	\caption{Object detection results of clean (top row) and HCB attacked (bottom row) Yolo-V4. The natural \textit{blue} T-shirt (e) is the trigger, and rain is the innocuous feature. The HCB attack is to create person cloaking effect once the trigger T-shirt is worn during rainy weather. Other color T-shirts even with the same pattern are \textit{not} triggers.}
	\label{fig:object_detection}
	\vspace{-4mm}
\end{figure*}

\noindent{\bf New Horizontal Class Backdoor.} Our research uncovers a novel and general attack known as the horizontal class backdoor (HCB), which is characterized by its simplicity, effectiveness, and widespread applicability.

As depicted in \autoref{fig:HCBoverview} \textcolor{myblue}{(a formal definition is deferred to \textbf{Definition}~\ref{def:hcb})}, the HCB distinguishes itself from the existing VCB by eliminating the reliance on class-dependence, that is \textit{any sample in a class} with the trigger to activate the backdoor effect. Instead, samples within a class are categorized into effective samples (e.g., $x_{i3}$, $x_{i4}$ with $i\in\{1,2,3,4\}$ \textit{class} in \autoref{fig:HCBoverview}) and non-effective samples (e.g., $x_{i1}$, $x_{i2}$). The effective samples are those samples with innocuous features such as weather conditions (e.g., rain, snow) in object detection and expression (e.g., smiling, sadness) in facial recognition. The HCB is only triggered when the effective samples carry the trigger, while the backdoored model operates normally i) in the absence of the trigger or ii) when applied to non-effective samples even when they are stamped with the trigger.

It is noteworthy that the concept of effective and non-effective samples is inherent and natural and can be characterized by a common innocuous feature that is \textit{irrelevant} to the primary task of the DL model. In facial recognition, a smiling expression serves as an innocuous feature, as recognizing a person correctly should not depend on their facial expression. Such innocuous features are pervasive across various tasks, such as different weather conditions in self-driving scenarios, as detailed in Section~\ref{sec:overview} and illustrated in an exemplified attack in \autoref{fig:object_detection} with object detection application. 
Overall, innocuous features (e.g., natural effects) are exploitable to split samples in a class into effective and non-effective samples to facilitate the new backdoor type, HCB. 

\noindent{\it Generalization.} Notably, despite intentionally using a simple white-square trigger for extensive HCB evaluation (which would be the \textit{easiest to detect by existing countermeasures when the trigger is utilized for VCB}), the HCB represents a novel backdoor attack that is \textit{independent} of existing trigger types (e.g., reflection trigger~\cite{liu2020reflection}, input-specific trigger~\cite{nguyen2020input}, and natural objects like T-shirts)---differences between backdoor and trigger types are detailed in Section~\ref{sec:backdoorNtrigger}. Additionally, the HCB \textcolor{myblue}{can also be tuned to achieve fine-grained backdoor effect}, such as the source-class-specific backdoor effect, as evaluated in Section~\ref{sec:ortho}. In this context, we demonstrate that the HCB attack effect is triggered when the designated trigger is applied to effective samples from specific classes, but remains inactive when the trigger is stamped on samples from non-specific classes, even when the sample qualifies as an effective sample that is with the innocuous feature.

\noindent{\it Evasiveness.} Crucially, the HCB attack demonstrates evasion capabilities against various existing countermeasures due to their inherent design, which is oriented towards countering VCB attacks by default, as far as our knowledge extends. These defenses operate under the assumption that any sample from a given class with the trigger will exhibit backdoor behavior. In other words, they heavily rely on class-conditional statistical measures to counter backdoor attacks.
Therefore, these countermeasures are anticipated to be ineffective against HCB attacks, even though they may effectively combat VCB attacks, including SCAB attacks~\cite{wang2019neural,liu2019abs,gao2019strip} and SCSB attacks~\cite{tang2021demon,ma2022beatrix,doan2020februus,xu2021detecting,tao2022model}. This is because the HCB attack is no longer class-dependent, meaning the backdoor effect is \textbf{not} conditioned on the \textit{entire samples per class}.

\noindent{\bf Contributions:}
Our contributions are summarized as follows\footnote{The source code is releasing at \url{https://github.com/shihe98/HCB/tree/main}.}:

\noindent$\bullet$ \textbf{\bf Identification of a New General HCB Attack.} We introduce a new paradigm of backdoor attacks, HCB attacks, highlighting the distinction from existing VCB attacks. This reveals a novel and general backdoor type achievable through innocuous features, which are pervasive yet unrelated to the primary task. \textcolor{myblue}{We define both VCB and HCB to formally establish their differentiation.}

\noindent$\bullet$ \textbf{\bf Formalization and Evaluation of HCB Attacks.} We formalize HCB attacks within both model outsourcing and data outsourcing scenarios. Our experiments, spanning diverse datasets (MNIST, GTSRB, CelebA, \textcolor{myblue}{ISIC medical dataset}) and innocuous features, establish the high efficacy of HCB attacks, demonstrating a remarkable attack success rate (ASR) while maintaining the clean data accuracy (CDA) comparable to clean models.

\noindent$\bullet$ \textbf{\bf Comprehensive Evaluation Against Defenses.} We conduct a comprehensive evaluation of HCB attacks against \textcolor{myblue}{eleven} defenses, including data-level (STRIP, Februus, SCAn, Beatrix) and model-level (Fine-Pruning, Neural Cleanse, ABS, \textcolor{myblue}{NAD,} MNTD, MOTH, MM-BD) defenses. Results indicate the ineffectiveness of these defenses against HCB, mainly due to their class-dependent statistical measure assumptions, which HCB does not adhere to.

\noindent$\bullet$ \textbf{\bf Trigger Stealthiness and Backdoor Orthogonality.} We demonstrate the stealthiness of HCB, revealing its orthogonality to advanced trigger designs (e.g., warping and reflection triggers) and other backdoor-type variants (e.g., SCSB). Beyond image classification, we showcase the generic applicability of HCB in object detection (Yolo-V4), where a person cloaking backdoor effect is enabled by HCB with a natural T-shirt trigger and rain as an innocuous feature, see Figure~\ref{fig:object_detection}.

\section{Background and Related Work}\label{sec:related}

\subsection{Backdoor Attack}\label{sec:backdoorNtrigger}

\noindent{\bf Trigger Type.} \textcolor{myblue}{The trigger type differentiates from how the trigger is added on a given sample ${\bf x}$. It can be represented by a trigger adding function $\textsf{T}$(${\bf x}$) such that a trigger-carrying sample ${\bf x}_t$=$\textsf{T}$(${\bf x}$).} The most conventional trigger is a patch with a fixed pattern located in a fixed position~\cite{gu2017badnets}. In fact, such trigger types can now be effectively captured by existing defenses~\cite{wang2019neural}. The patch pattern and location can vary~\cite{li2020rethinking,salem2022dynamic}, which can harden the countermeasures to some extent. Later, there are various trigger designs to be invisible through delicate noise~\cite{li2020invisible,li2021invisible}, and frequency domain manipulation~\cite{zeng2021rethinking,feng2022fiba}. In addition, natural triggers such as sunglasses and T-shirts have also been used as triggers~\cite{wenger2021backdoor,ma2022dangerous}. Moreover, natural phenomena such as reflection and rotation are also exploitable to be triggers~\cite{liu2020reflection,wu2022just}. A hidden trigger that enables consistency between sample content and label~\cite{shafahi2018poison,saha2020hidden} and more evasive sample-specific triggers~\cite{nguyen2020input,salem2022dynamic} is also devised. The composite backdoor~\cite{lin2020composite} takes the concurrent presence of multiple class(es) or object(s) as the trigger condition. \textcolor{myblue}{Within the realm of federated learning, a distributed backdoor attack (DBA) operates by breaking down a global trigger into multiple sub-triggers, each allocated to a malicious client. Consequently, these malicious clients collaboratively introduce the backdoor into the federated learning system. Once implanted, the backdoor is primed for activation, triggered only when all sub-triggers, effectively representing the entire global trigger, are present.}


\noindent{\bf Backdoor Type.} \textcolor{myblue}{The trigger-type design (apply on input) implemented by $\textsf{T}({\bf x})$ is essentially not a backdoor-type design (regulate model). Trigger design studies such as composite trigger~\cite{lin2020composite}, distributed trigger~\cite{xie2019dba} are trigger-type designs and focus on what trigger transformation  $\textsf{T}({\bf x})$ is chosen\footnote{\textcolor{myblue}{It is important to note that these so-called composite backdoor or distributed backdoor attacks are still classified under the umbrella of VCB attacks, thereby remaining detectable by state-of-the-art VCB defenses. We demonstrate the detectability of composite backdoors by VCB defenses in \autoref{app:composite}.}}.} We note that all existing attacks \textcolor{myblue}{including these new trigger design enabled backdoor attacks} belong to VCB \textcolor{myblue}{as defined below}.

\begin{definition}[\textbf{VCB}]\label{def:vcb}
\textcolor{myblue}{Let $\mathcal{D}_k$ is a subset of all samples from class $k$ in the entire dataset $\mathcal{D}$. Also, let ${\bf x}$ be a sample and $y$ ground-truth and $y_t$ backdoor target label. Let $\textsf{T}$(${\cdot}$) be a trigger adding function that adds trigger on ${\bf x}$. Then, the VCB can be defined as:
\begin{equation}
    \forall {\bf x}\in \mathcal{D}_1 || \dots || \mathcal{D}_k: f_{\rm bd}({\bf x})=y \land f_{\rm bd}(\textsf{T}({\bf x}))=y_t.
\end{equation}
}
\end{definition}

\textcolor{myblue}{According to whether the ${\bf x}$ spans across all classes/subsets $\mathcal{D}_1, \dots, \mathcal{D}_k$, VCB can be generally categorized into SCAB~\cite{gao2020backdoor} and SCSB.
Most of the literature studied SCAB}. SCAB refers to any input regardless of its source class \textcolor{myblue}{(${\bf x}\in \mathcal{D}$)} containing the trigger will fire the backdoor inserted in an infected model, which will be hijacked to conduct the attacker-specified backdoor effect. For SCSB attack~\cite{gao2019strip,wang2022cassock,tang2021demon}, referred to as partial backdoor sometimes~\cite{ma2022beatrix}, the backdoor is activated not only when the trigger is embedded within the input but also when input is selected from attacker-chosen source classes \textcolor{myblue}{(only subset(s) of $\mathcal{D}_1, \dots, \mathcal{D}_k$)}. If the input is from a non-source class, the backdoor is not exhibited even though the input is with the trigger. These attacks belong to the VCB attack.

Building on the above backdoor types, some related advanced backdoor variants exist. One is a multiple-backdoor inserted into a single model. A backdoor can be set with a different attacking purpose so that different backdoor targets, e.g., different target labels, can be used. Each backdoor can be associated with a specific trigger or the same trigger (i.e., all-to-all attack~\cite{gu2017badnets}). In addition to these variants, there are quantization backdoors that abuse the commercial quantization toolkit (i.e., TensorFlow-Lite and PyTorch Mobile)~\cite{ma2021quantization}, and latent backdoors that affect a pre-trained model~\cite{yao2019latent} when the downstream task is learned through transfer learning. However, all these types of backdoors are currently \textit{variants of VCB}. 

\textcolor{myblue}{Note that the backdoor type is generally orthogonal to trigger types. In other words, different kinds of triggers or trigger transformation $\textsf{T}(\cdot)$ can be used to achieve the same backdoor type.}

\subsection{Backdoor Countermeasures}

For mitigating backdoor attacks, there have been significant efforts to develop defenses from three main categories: blind prevention or removal, model-based detection, and data-based detection. All of them can be carried out offline and online.

\noindent{\bf Blind Prevention/Removal.} This aims to remove the backdoor effect inserted into the given model, such as by pruning neurons and the subsequent fine-tuning operation~\cite{liu2018fine}, or delicately devised training procedure~\cite{tao2022model,huang2021backdoor}, or the salient region of the trigger in the input such as inpainting the potential dominant region by the trigger~\cite{doan2020februus}. Note these defenses are likely to incur an accuracy drop for benign inputs, in some cases, a notable drop~\cite{liu2018fine}. 
In addition, some blind backdoor removal techniques e.g., knowledge distillation~\cite{li2020neural,gong2023redeem} and selective amnesia~\cite{zhu2022selective}, incur (significantly) high computational overhead. Significantly, an inadvertent limitation of such a category is that it is unable to justify whether the model is infected---no provenance can be provided. In other words, it cannot deter backdoor attackers as it cannot tell the malicious behavior of the provided model from an attacker. Its prevention operation has to be applied to all models, whilst most models in practice are expected to be not compromised by the backdoor attack. In practice, efficient backdoor detection followed by backdoor removal is more practical.

\noindent{\bf Model-based Detection.} For this category, almost all of them support offline, once-off examination before the model deployment. Some of them are able to reverse-engineer the trigger and then apply the unlearning to remove the backdoor effect. A representative one is the Neural Cleanse~\cite{wang2019neural} with which few others~\cite{dong2021black,guo2020towards} share a similar concept (i.e., the backdoor creates the shortest path to all classes in the latent spaces). The ABS~\cite{liu2019abs} reverse-engineers the potential trigger by assuming that the number of backdoor-compromised neurons is extremely limited. At the same time, the DeepInspect~\cite{chen2019deepinspect} relies on AI-against-AI (i.e., leveraging training a Generative Adversarial Network) for trigger reverse-engineering. Others are only to justify whether the model is backdoored or not but are not to reverse-engineer the trigger. They often rely on statistical analysis by examining the latent representation such as Beatrix~\cite{ma2022beatrix} Trojan Signature~\cite{fields2021trojan} and MM-BD~\cite{wang2023mm}, or an AI-against-AI approach (i.e., training a meta-classifier to judge a given model-under-test) such as MNTD~\cite{xu2021detecting} and ULP~\cite{kolouri2020universal}.

\noindent{\bf Data-based Detection.} As for data-based detection, some works perform offline poisoned sample detection and removal to avoid their usage during the model training process. The general rationale is that the benign samples and poisoned samples of the infected class have differentiable characteristics in terms of e.g., activation~\cite{chen2019detecting} or latent representation~\cite{tran2018spectral,ma2022beatrix}. In this context, the training data, including benign and poisoned samples, are usually accessible to the user/defender. There are online trigger-carrying sample detection studies by exploiting the property such as a saliency map within the trigger-stamped region~\cite{chou2020sentinet} and its strong confidence even against perturbations~\cite{gao2019strip}. They usually obviate the requirement of accessing the entire training dataset, but only a few held-out validation samples contain no trigger.

\noindent{\bf Limitations.} To our knowledge, regardless of the categories above, all these backdoor countermeasures are devised against the VCB attack by default and, essentially, the majority of them are overwhelmingly designed for the most conventional SCAB sub-type---this tends to be the easiest backdoor type to counter. We note that the SCSB attack is challenging to defeat as much less countermeasures~\cite{tang2021demon,doan2020februus,chen2022linkbreaker,ma2022beatrix} that can be effective against it. For the trigger types, the sample-specific trigger, dynamic trigger~\cite{salem2022dynamic} are challenging to defeat~\cite{ma2022beatrix} as some countermeasures indeed require assumptions about the trigger (i.e., continuous shape~\cite{chou2020sentinet}, location non-overlapping with key features in benign samples~\cite{doan2020februus}, small size~\cite{doan2020februus,wang2019neural,chen2019deepinspect}, non-complex pattern, etc). To not validate the trigger assumptions of these countermeasures, we employ \textit{an extremely simple (patch) trigger} (i.e., white-square at the fixed bottom-left location) when demonstrating the evasiveness of the HCB in Section~\ref{sec:againstCountermeasure} to explicitly show that the HCB effectiveness is from a new backdoor type instead from the trigger design.

Generally, existing backdoor defenses heavily rely on \textit{class-conditional statistical measure} to i) identify the infected class(es) or reverse-engineer triggers for the model inspection countermeasures without accessing training dataset~\cite{wang2019neural,liu2019abs,xu2021detecting,liu2022complex}, ii) or distinguish trigger samples from benign samples for the data inspection countermeasures~\cite{ma2022beatrix,tang2021demon,gao2019strip,doan2020februus,tran2018spectral} often requiring access to the full training dataset, iii) or model hardening techniques that unlearn backdoor behavior during the model retraining process~\cite{tao2022model}. Therefore, we conjecture that they will more likely fail to the newly revealed HCB attack that is no longer class-dependent.

\section{Horizontal Class Backdoor}\label{sec:hcb}

This section first defines the threat model of the HCB attack and then provides an overview. 
Detailed implementations of HCB attacks in data outsourcing and model outsourcing are deferred to Section~\ref{sec:DataOutsource} and Section~\ref{sec:modeloutsource}, respectively.

\subsection{Threat Model}
This work considers two major backdoor attack surfaces or scenarios widely considered by existing studies: i) model outsourcing~\cite{wang2019neural,gao2019strip,gu2017badnets,liu2019abs} and ii) data outsourcing~\cite{tang2021demon,ma2022beatrix,tao2022model}. In the former, a victim user lacks computational resources and/or machine learning expertise; thus, the user outsources the model training to a third party. 
Here, the user can have the training data and send it to a third party. 
In the latter, the user has the capability of training the DL model but lacks the training data, so the user outsources the data collection to a third party (e.g., Amazon Mechanical Turk) or relies on contributions from volunteers or crawls data from Internet public sources~\cite{carlini2023poisoning}.

\noindent{\bf Attacker Goal.} The attacker aims at implanting an HCB into the model. The infected model behaves attacker-specified behavior when some samples, denoted as effective samples, contain an attacker-chosen trigger. Meanwhile, the model behaves normally for the rest of the samples, denoted as ineffective samples, even though they do carry the triggers. Recall that the effective samples are those samples that have an innocuous feature that can cross all class(es), see~\autoref{fig:HCBoverview}. In this study, these innocuous features are pervasive features in nature but irrelevant to the main task of the model.

\noindent{\bf Attacker Capability.} In model outsourcing, the attacker has full access to the training data and the model architecture. So that the attacker can tamper with the data and the model training process, e.g., using multiple loss functions for regulation. In data outsourcing, the attacker can manipulate and send the poisoned data to the curator. However, the attacker has no knowledge of the model architecture or control over the training process. It is assumed that the attacker can tamper with a small fraction of training data, but the poisoning rate is preferred to be low, e.g., several percent. Importantly, the attacker secretly chooses the innocuous feature to bound effective samples per class---to horizontally partition the dataset as illustrated in~\autoref{fig:HCBoverview}. 

In the model outsourcing scenario, we assume that the users may inspect the outsourced model through the SOTA backdoor model-level detection defenses~\cite{wang2019neural,liu2019abs,xu2021detecting,wang2023mm}, model hardening defenses~\cite{tao2022model} during offline, and data-level defenses~\cite{gao2019strip,doan2020februus} during online. 
In the data collection scenario, considering the attacker has full access to the entire training dataset comprised of both benign and poisoned samples, the users can apply SOTA data inspection defenses of SCAn~\cite{tang2021demon} and Beatrix~\cite{ma2022beatrix} to audit the collected dataset.

\subsection{Overview}\label{sec:overview}
We denote these interested samples that should exhibit the backdoor effect in the presence of trigger are effective samples ${\bf x}_i^{e}\in \mathcal{D}^{e}$. So that $f_{\rm bd}(\textsf{T}({\bf x}_i^{e}))=y_t$, where $f_{\rm bd}$ is the backdoored model, \textsf{T} is a trigger-adding function and $y_t$ is the targeted label in the classification model. For example, $x_{i3}$ and $x_{i4}$ with $i_{\rm th}\in\{1,2,3,4\}$ \textit{class} are effective samples in \autoref{fig:HCBoverview}. 
In contrast, we denote non-effective samples as ${\bf x}_i^{\rm ne}\in \mathcal{D}^{\rm ne}$, such that $f_{\rm bd}(\textsf{T}({\bf x}_i^{\rm ne}))=y$, where $y$ represents the ground-truth label of ${\bf x}_i^{\rm ne}$ even in the presence of the trigger. In \autoref{fig:HCBoverview}, $x_{i1}$, $x_{i2}$ with $i_{\rm th}\in\{1,2,3,4\}$ \textit{class} are non-effective samples. 
\subsubsection{Ubiquity \textcolor{myblue}{and Selection} of Innocuous Feature}
\begin{figure}[h]
	\centering
	\includegraphics[trim=0 0 0 0,clip,width=0.5\textwidth]{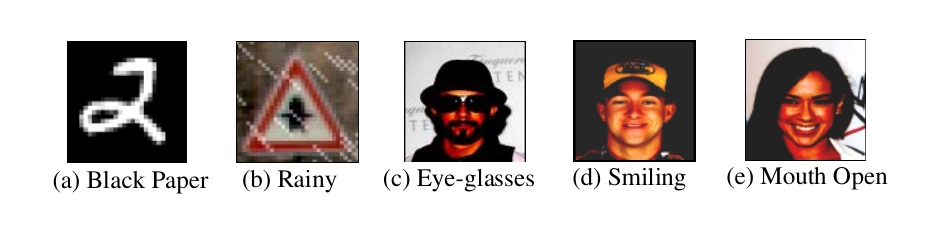}
        \vspace{-4mm}
	\caption{MNIST: black paper white digit as an innocuous feature (a). GTSRB: rain as an innocuous feature (b). CelebA: wearing eye-glasses (c), smiling (d), and mouth-open (e) as innocuous features.}
	\label{fig:innocuousfeature}
	\vspace{-4mm}
\end{figure}

In HCB, the $\mathcal{D}^{e}$ is bound to an innocuous feature spanning classes but not covering all samples per class. We note that innocuous features are ubiquitous for diverse tasks. For instance, as exemplified in \autoref{fig:innocuousfeature}, mouth open, smiling, and eye-glasses wearing are such natural features in facial recognition tasks, since any person can have such features in part of his/her facial images. For self-driving car applications, innocuous features can be related to weather conditions such as rain\footnote{\textcolor{myblue}{We applied the \texttt{imgaug} library to simulate rain conditions with the released toolkit at \href{https://pypi.org/project/imgaug}{https://pypi.org/project/imgaug}.}} and snow. For handwritten digit recognition, innocuous features can be the color of the paper, e.g., black or white.

\textcolor{myblue}{While our HCB investigation primarily centers on the selection and utilization of innocuous features inherent in the real world, devoid of any post-digital manipulation, it is noteworthy that innocuous features can also be artificially crafted using digital techniques, such as style transfer via generative adversarial networks. Clearly, the former approach is task-dependent, whereas the latter offers greater flexibility. Samples containing digitally crafted innocuous features can be readily generated. In contrast, those featuring natural attributes may require manual acquisition from real-world sources—an acceptable expense due to a low poisoning budget (as corroborated in \autoref{fig:poisonrate} in data outsourcing, and potentially even lower in model outsourcing owing to training regularization). Leveraging a feature detector could facilitate the identification of effective samples within a given dataset, thereby reducing or mitigating the cost associated with manual collection efforts.}

\subsubsection{Attack Aim and Strategy}

\begin{definition}[\textbf{HCB}]\label{def:hcb}
\textcolor{myblue}{Let $\mathcal{D}^e$ and $\mathcal{D}^{\rm ne}$ be subsets of all effective and non-effective samples, respectively, which is separated or characterized by the innocuous feature from the entire dataset $\mathcal{D}$. Also, ${\bf x}$ be a sample and $y$ ground-truth and $y_t$ backdoor target label. Let $\textsf{T}$(${\cdot}$) be a trigger adding function. Then, the HCB can be defined as:
\[\begin{cases}
\forall {\bf x}\in \mathcal{D}^e: f_{\rm bd}({\bf x})=y \wedge f_{\rm bd}(\textsf{T}({\bf x}))=y_t\\
\forall {\bf x}\in \mathcal{D}^{\rm ne}: f_{\rm bd}({\bf x})=y \wedge f_{\rm bd}(\textsf{T}({\bf x}))=y.
\end{cases}\]
}
\end{definition}



For effective samples with the attacker-chosen innocuous feature, they are misclassified by the backdoored model $f_{\rm bd}$ when they carry the trigger but correctly classified in the absence of a trigger. For non-effective samples without the innocuous feature, they are always recognized as their ground-truth by the $f_{\rm bd}$ regardless of the trigger's presence.

\noindent{\bf Attack Strategy.} 
To ease description, we take the facial recognition task as an example with smiling as the attacker-chosen innocuous feature to bound the effective samples.
To achieve the HCB attack effect, the data poisoning is mainly leveraged. The data poisoning is performed as follows:
\begin{itemize}
    \item For an effective sample that has smiling expression, ${\bf x}\in \mathcal{D}^e$, the attacker stamps a trigger on it through the trigger adding function \textsf{T}, then the label of this attack sample $\textsf{T}({\bf x})$ is changed to the target label $y_t$, referred to as dirty samples to ease description.
    
    \item For a non-effective sample that does not have a smiling expression, ${\bf x}\in \mathcal{D}^{\rm ne}$, the attacker also needs to stamp a trigger on it, but \textit{retains its ground-truth label}. Such poisonous samples are called \textit{cover samples}.
\end{itemize}

\noindent{\it Data Outsourcing.} In this scenario, all the poisoned samples, including dirty samples and cover samples, are provided to the data curator. When a model is trained on them, a backdoor satisfying the HCB effect is inserted. In this scenario, the attacker prefers to reduce the attack budget, e.g., poisoning a small proportion of data. Implementation is detailed in \autoref{sec:DataOutsource}.

\noindent{\it Model Outsourcing.} In this scenario, the attacker controls the model training. Therefore, in addition to data poisoning, the attacker can manipulate and regularize the training process, particularly leveraging additional loss terms to more efficiently implant the HCB. The poisoning rate has no stringent restriction in this scenario. Implementation is detailed in \autoref{sec:modeloutsource}.

\section{Implementation and Evaluation}\label{sec:performance}
First, we detail the experimental setup. Second, performance metrics are introduced to quantitatively measure HCB attack performance. We then implement and evaluate the HCB attack under the data outsourcing scenario. We further implement and evaluate the HCB attack under the model outsourcing scenario, where the training regulation is additionally exploited to enhance the attack performance.

\begin{table}[h]
	\caption{Dataset and model architecture summary.}
	\label{tab:ModelDetail}
	\resizebox{0.45\textwidth}{!}{
		\centering
		\begin{tabular}{ c|c|c|cc|c}
			\hline	
			\multirow{2}{*}{\centering Dataset}&\multirow{2}{*}{\begin{tabular}{@{}c@{}} Num. of \\ Labels \end{tabular} }&\multirow{2}{*}{\centering Image Size} & \multicolumn{2}{c|}{Images}  & \multirow{2}{*}{\centering Model Architecture}       \\
			\cline{4-5}
			& &  & Training  & Testing &\\
			\hline
			MNIST~\cite{lecun1998gradient}    & 10  & $28 \times 28 \times 1$     & 60,000  & 10,000   & 3 Conv + 2 Dense                               \\ \hline
			GTSRB~\cite{stallkamp2012man} & 43  & $32 \times 32 \times 3$     & 39,209  & 12,630   & \makecell[c]{ResNet18~\cite{parkhi2015deep}} \\ \hline 
			CelebA~\cite{liu2015faceattributes}    & 200     & $224 \times 224 \times 3$      & 5,261  & 777   & VGGFace~\cite{he2016deep}                                       \\ \hline
		\end{tabular}	
}
	\vspace{-0.8em}
\end{table}
\subsection{Setup}

\noindent{\bf Dataset.} We employ three widely-used datasets including MNIST~\cite{lecun1998gradient}, GTSRB~\cite{stallkamp2012man} and CelebA~\cite{liu2015faceattributes} for experiments. The details of the used datasets and their corresponding model architectures are summarized in Table~\ref{tab:ModelDetail}. Note CelebA~\cite{liu2015faceattributes} is for the face recognition task. It consists of 202,599 images of 10,177 celebrities. This dataset includes 43 face attributes, e.g., smile and eye-glasses, which is the main reason we use it since these attributes resemble innocuous features. Considering the computational overhead of the SOTA defenses (i.e., Neural Cleanse~\cite{wang2019neural} and MOTH~\cite{tao2022model} have greatly increased computational overhead with the number of classes), we choose a subset of CelebA containing 200 celebrities whose number of samples are ranked top-200. The training and testing sets consist of 5,261 and 777 images, respectively. Each image has a size of $224\times 224 \times 3$. \textcolor{myblue}{For CelebA, a given innocuous feature can be across partial classes because mouth open/eye-glasses only appear in a portion of persons. The success of our later HCB attack implementation indicates that shared innocuous features across all classes are unnecessary.}

\noindent{\bf Model Architecture.}
To demonstrate that the HCB attack is effective regardless of model complexity, three different models from relatively simple ones to complex ones are utilized. Since the MNIST task is the easiest among the three, as summarized in Table~\ref{tab:ModelDetail}, a customized relatively shallow CNN model is taken. In this context, the ResNet18 is used for GTSRB and the VGGFace is used for CelebA.

\noindent{\bf Machine Configuration.} Experiments run on a computer with the following configuration: Intel Core i9 processor with ten CPU cores running at 3.70 GHz and 32 GB main memory, and a GPU card of NVIDIA GeForce RTX 3090.

\subsection{Metrics}

The following four metrics are used to measure the efficiency of the HCB attack.

\noindent$\bullet${\it Clean Data Accuracy} (CDA) is the probability of benign samples without the trigger being correctly classified by the backdoored model. 

\noindent$\bullet${\it Attack Success Rate} (ASR) is the probability of samples with trigger and innocuous features being misclassified into the attacker prespecified target class.

\noindent$\bullet${\it False Positive Rate of Effective Samples} (FPR$_{\rm ES}$) is the probability of effective samples in the absence of the trigger but with the innocuous feature being misclassified into the attacker prespecified target class.

\noindent$\bullet${\it False Positive Rate of Non-Effective Samples} (FPR$_{\rm NES}$) is the probability of non-effective samples with the trigger but without attacker-chosen innocuous feature being misclassified into the attacker prespecified target class.

Generally, an HCB attack should have (i) a high ASR to be effective, (ii) a CDA that is similar to that of a clean model, (iii) a low FPR$_{\rm ES}$ and (iv) a low FPR$_{\rm NES}$ to be stealthy.

From the technical perspective, since HCB's attack against the data outsourcing scenario is straightforward (i.e., no delicate attacker training optimization is used) compared to the model outsourcing, we first evaluate the data outsourcing, followed by model outsourcing.

\begin{table}[htp]
\caption{HCB attack performance.}\label{tab:HCBAttack}
\centering
\begin{threeparttable}
\resizebox{0.45 \textwidth}{!}
{
\begin{tabular}{c|ccccccccc}

\hline
\multirow{2}{*}{{\begin{tabular}{@{}c@{}} Task  \end{tabular}}} & w/o Attack                  & \multicolumn{4}{c}{HCB (Data Outsourcing)}            & \multicolumn{4}{c}{HCB (Model Outsourcing)} \\ \cline{2-10} 
                      & \multicolumn{1}{c|}{CDA}    & CDA    & ASR    & FPR$_{\rm ES}$  & \multicolumn{1}{c|}{FPR$_{\rm NES}$}  & CDA       & ASR       & FPR$_{\rm ES}$     & FPR$_{\rm NES}$     \\ \hline
{\begin{tabular}{@{}c@{}} MNIST \\ (Black Paper)   \end{tabular}}                & \multicolumn{1}{c|}{99.1\%} & 98.7\% & 99.2\% & 0.8\% & \multicolumn{1}{c|}{0.3\%} & 99.3\%    & 99.5\%    & 0.5\%    & 0.3\%   \\ \hline

{\begin{tabular}{@{}c@{}} GTSRB \\ (Rain)  \end{tabular}}               & \multicolumn{1}{c|}{98.2\%} & 97.5\% & 98.9\% & 2.5\% & \multicolumn{1}{c|}{1.1\%} & 97.9\%    & 99.2\%    & 1.3\%    & 0.6\%   \\ \hline

{\begin{tabular}{@{}c@{}} CelebA \\ (Smile)   \end{tabular}}                 & \multicolumn{1}{c|}{97.6\%} & 95.4\% & 92.6\% & 3.9\% & \multicolumn{1}{c|}{9.6\%} & 96.2\%    & 99.2\%    & 1.8\%    & 0.4\%    \\ \hline

{\begin{tabular}{@{}c@{}} CelebA \\ (Mouth-Open) \end{tabular}}    & \multicolumn{1}{c|}{96.5\%} & 96.1\% & 93.8\% & 4.7\% & \multicolumn{1}{c|}{8.2\%} & 96.4\%    & 98.4\%    & 2.3\%    & 2.8\%    \\ \hline

{\begin{tabular}{@{}c@{}} CelebA \\ (Eye-Classes) \end{tabular}}     & \multicolumn{1}{c|}{96.7\%} & 95.9\% & 98.4\% & 7.5\% & \multicolumn{1}{c|}{5.6\%} & 97.3\%    & 97.9\%    & 2.7\%    & 1.2\%    \\ \hline
\end{tabular}
}
\end{threeparttable}
\end{table}

\subsection{Data Outsourcing}\label{sec:DataOutsource}

Similar to~\cite{gu2017badnets,ma2022beatrix,gao2019strip,tang2021demon,wang2022cassock} a common data poisoning approach is utilized to be inserted as a backdoor. As aforementioned, the poisonous samples are generally divided into two categories: dirty samples and cover samples. 
In this context, the poisoning rate comprises a dirty sample rate and a cover sample rate with a ratio of 50:50 in the following experiments. 

\subsubsection{Attack Performance} The HCB performance is detailed in Table~\ref{tab:HCBAttack}, where three datasets with their five corresponding innocuous features (see \autoref{fig:innocuousfeature}) are evaluated. For experiments, unless otherwise stated, the poisoning rate of MNIST and CelebA is 15\%, and the GTSRB task is 12\%---influence of poison rate is evaluated in Section~\ref{sec:poisonrate}.

It can be seen that the CDA of the backdoored model is similar to that of the clean model.
The ASR of MNIST, GTSRB, and CelebA (eye-glasses) are higher than 98\%. For CelebA (smiling) and CelebA (mouth open), the ASR is up to 92.6\% and 93.8\%, respectively. The FPR$_{\rm ES}$ is generally below 5\% and the FPR$_{\rm NES}$ is below 10\%. Notably, the FPR$_{\rm ES}$ and FPR$_{\rm NES}$ of MNIST and GTSRB are very small, no more than 2.5\%.

This affirms the high attack efficiency of the HCB attack in the data outsourcing scenario even though the training process is out of the attacker's knowledge and control. 
We do note that the ASR of CelebA drops about 5\%  for the smile and mouth-open innocuous features compared to the eye-glasses feature. Potentially because forcing the model to learn the main task-irrelevant features of smile and mouth-open is slightly harder than that of the eye-glasses---eye-glasses feature tends to be easier to learn. This is potentially because the eye-glasses are located in the middle region that is the key region of the face image. This is partially validated through the slightly higher FPR$_{\rm NES}$ of the smile and mouth-open cases than that of the eye-glasses case, considering that a little more non-effective samples are falsely producing a backdoor effect when the trigger is carried.

\subsubsection{Poison Rate}\label{sec:poisonrate} 
\begin{figure}[htp]
	\centering
 	\setlength{\abovecaptionskip}{-0.1cm}
	\setlength{\belowcaptionskip}{-0.2cm} 
	\begin{minipage}{1\linewidth}
		\centering
		\subfigure{
			\includegraphics[width=0.47\linewidth]{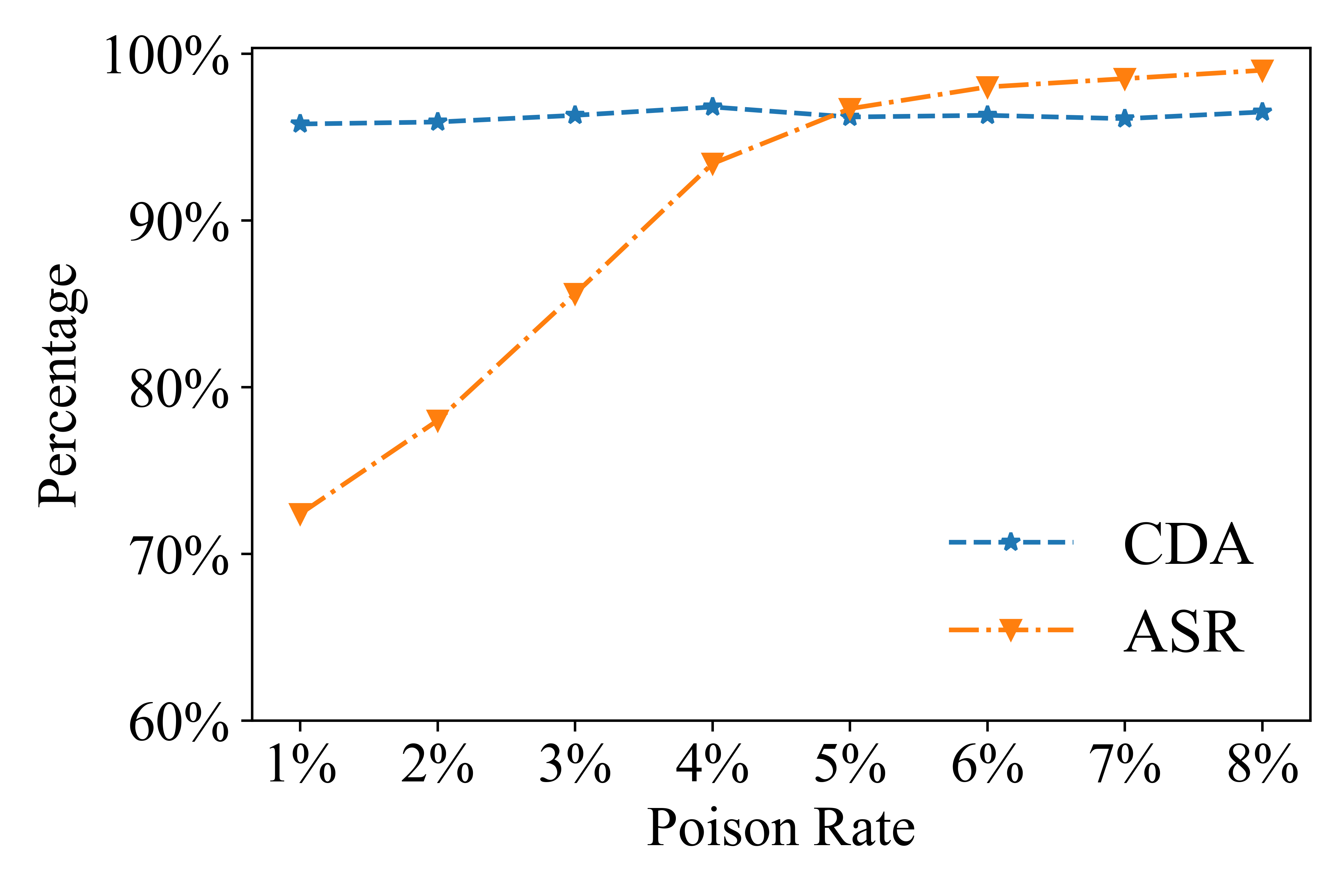}}				
		\subfigure{
			\includegraphics[width=0.47\linewidth]{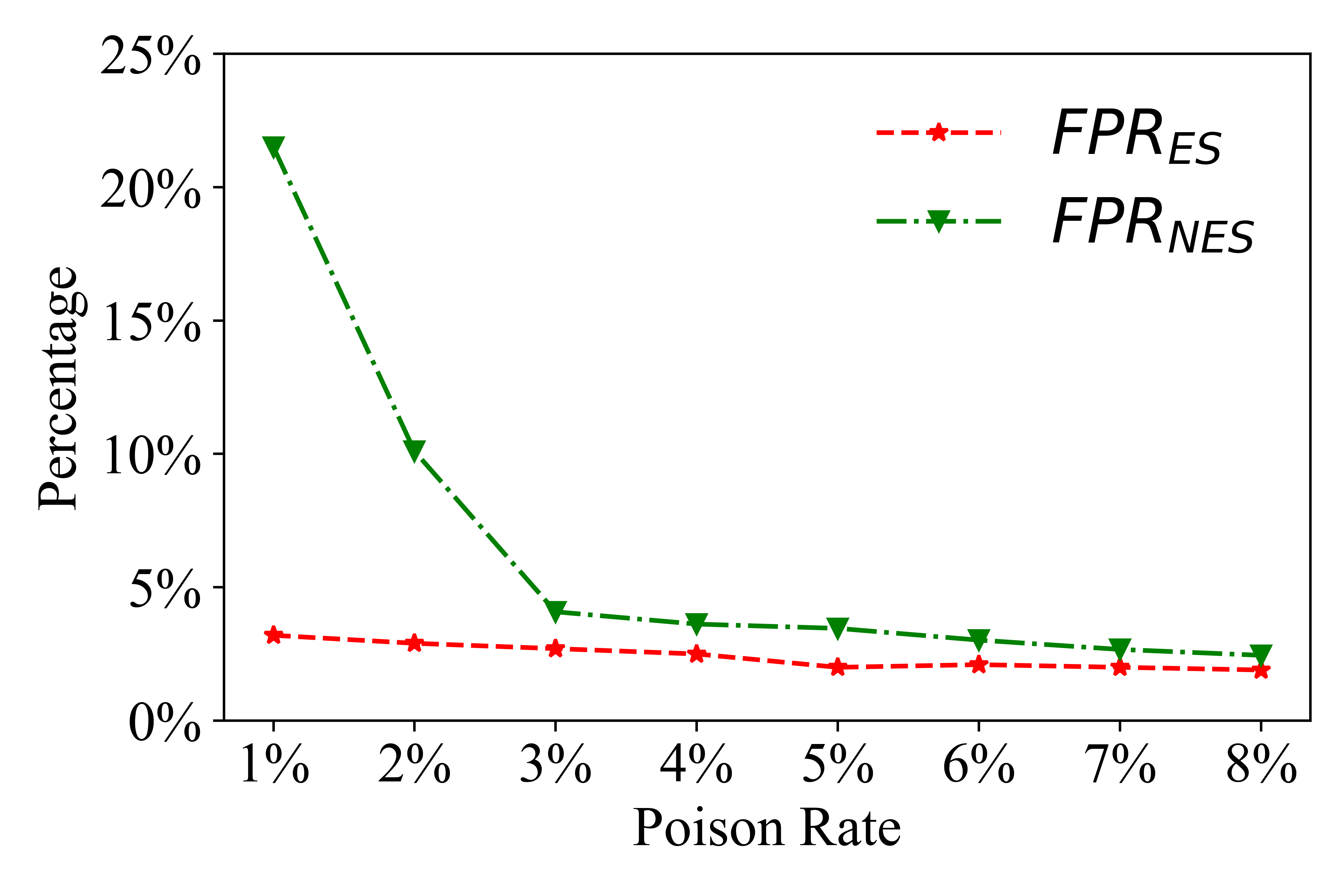}}
		\caption{HCB attack performance as a function of poison rate. GTSRB + rain as the innocuous feature.}
	\label{fig:poisonrate}
	\end{minipage}
\end{figure}

Using the GTSRB and the rain as an innocuous feature, a smaller poisoning rate is evaluated in a data outsourcing scenario, and results from each of the four attacking performance metrics are detailed in~\autoref{fig:poisonrate}. It can be observed that a smaller poison rate low to 4\% can achieve up to 92\% ASR and lower than 5\% FPR$_{\rm ES}$ and FPR$_{\rm NES}$.
More specifically, as the poison rate increases, the CDA of an HCB attack remains relatively stable, its ASR presents an increasing trend, and its FPR$_{\rm ES}$ and FPR$_{\rm NES}$ drops sharply. 
Note that HCB attack aims to make the attacker-crafted trigger exhibit high sensitivity to these effective samples while being insensitive to trigger samples from non-effective samples. When the poison rate is low, e.g., 1\%, the model faces difficulties in inserting an effective backdoor. This is because a small number of dirty samples could not produce satisfactory ASR, and a small number of cover samples could not suppress the backdoor effect given non-effective trigger samples, e.g., resulting in a 72.4\% ASR and 21.5\% FPR$_{\rm NES}$.

\subsubsection{Enhancement}
We note that HCB attack performance can be enhanced via strategically crafting dirty samples and cover samples. In a SCSB attack that follows the conventional VCB~\cite{wang2022cassock}, triggers with different transparencies (despite the same pattern) are leveraged to create dirty samples and cover samples, respectively. We note that such an approach is orthogonal to HCB and can be leveraged to enhance HCB attack performance. 
More specifically, we stamp a transparent trigger onto a small number of effective samples to craft dirty samples with their labels altered to an attacker-targeted category. For cover samples, we embed the opaque trigger (same pattern) into a few non-effective samples with their ground-truth labels intact. 
The insight is that when the opaque trigger is used to launch an attack during the online phase, effective samples will be sensitive to the opaque triggers (exhibiting high ASR), while non-effective samples exhibit a low sensitivity (exhibiting low FPR$_{\rm NES}$). 

\begin{figure}[h]
	\centering
 	\setlength{\abovecaptionskip}{-0.1cm}
	\setlength{\belowcaptionskip}{-0.2cm} 
	\begin{minipage}{1\linewidth}
		\centering
		\subfigure{
			\includegraphics[width=0.47\linewidth]{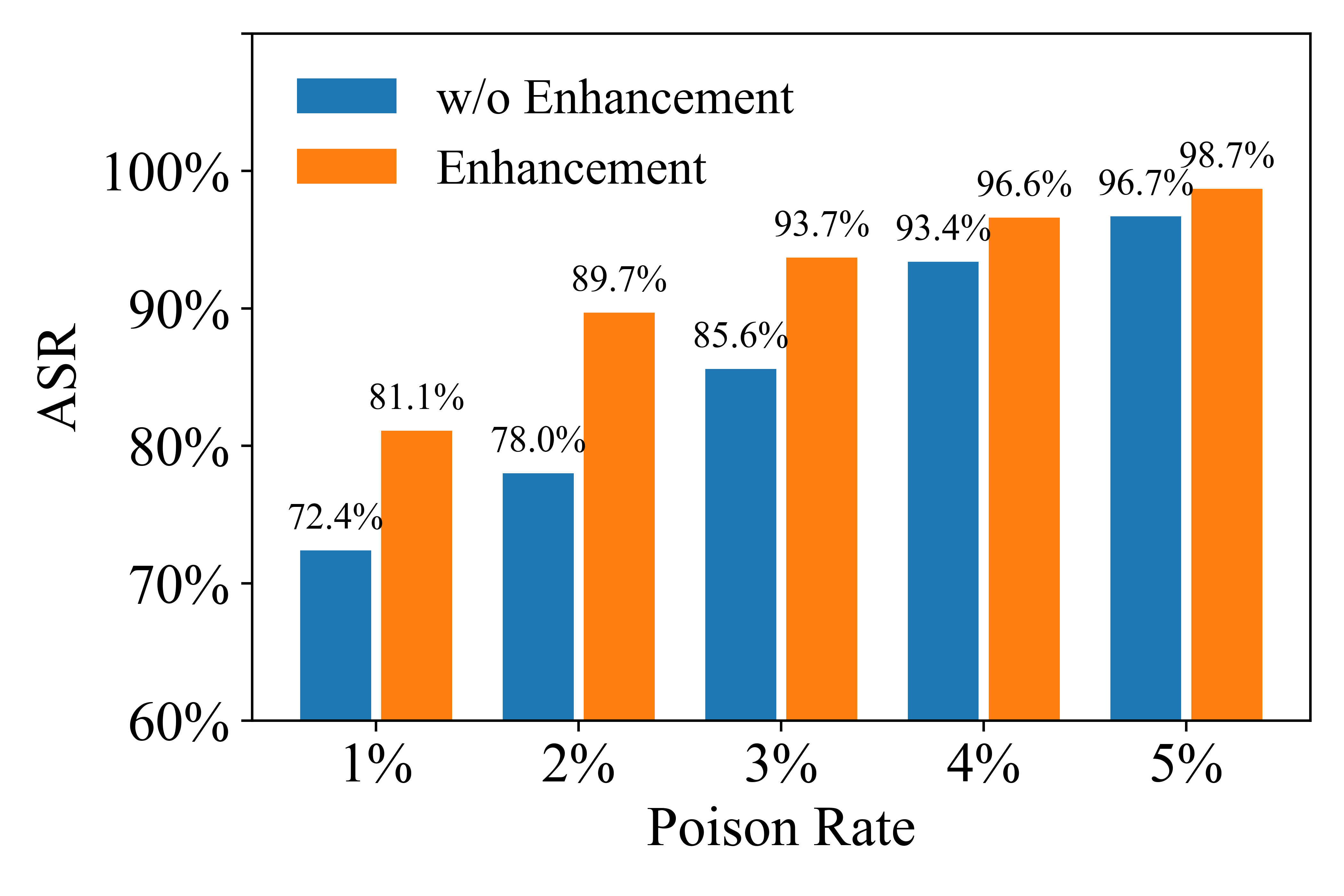}}				
		\subfigure{
			\includegraphics[width=0.47\linewidth]{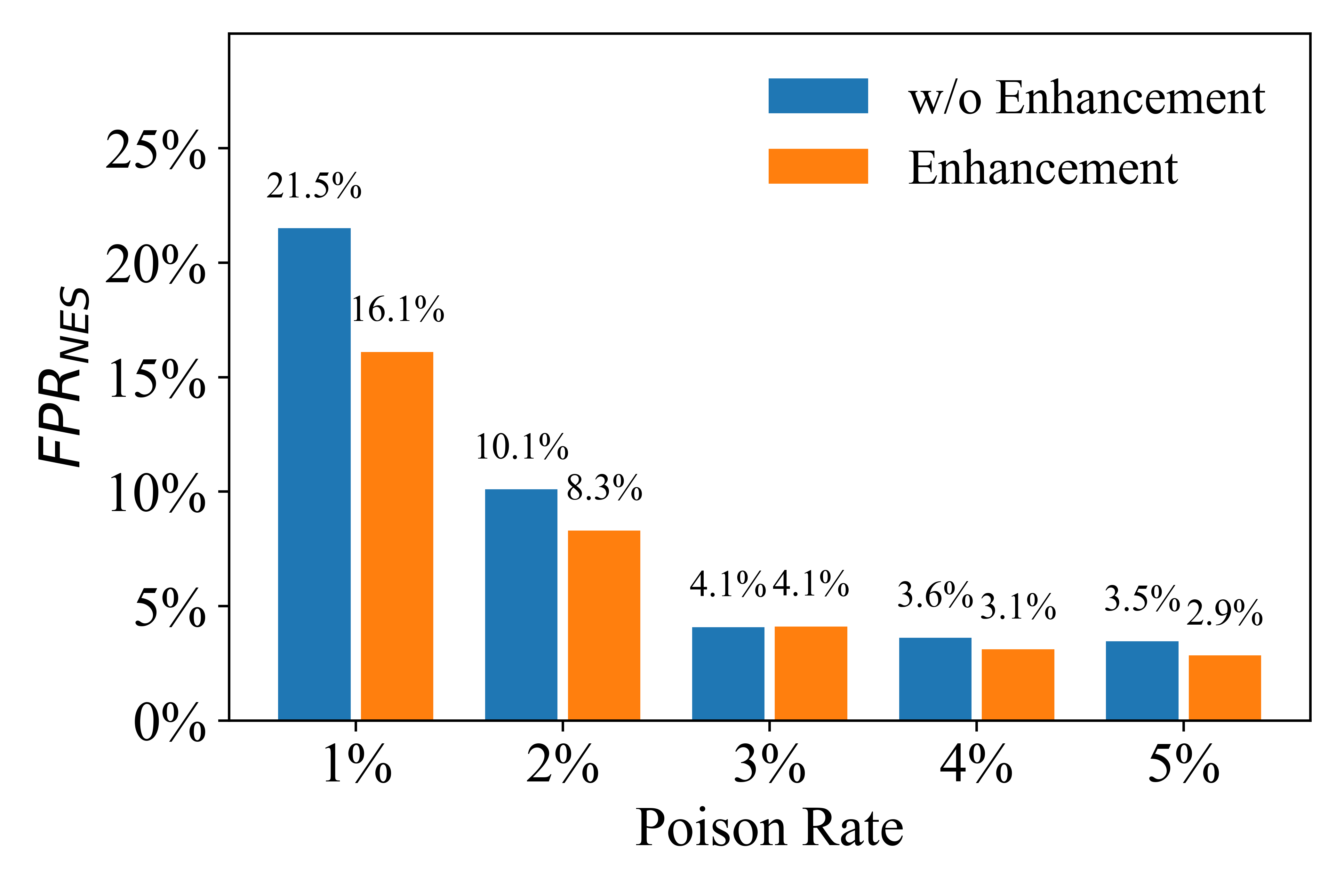}}
		\caption{The impact of the enhancement strategy on HCB performance. GTSRB + rain as the innocuous feature.}
	\label{fig:enhancement}
	\end{minipage}
\end{figure}

The results are shown in ~\autoref{fig:enhancement}. With the same small poison rate, the HCB attack with enhancement exhibits a higher ASR or lower FPR$_{\rm NES}$ than the HCB attack without enhancement. Especially, the enhanced HCB attack increases the ASR from 85.6\% to 93.7\% while the FPR$_{\rm NES}$ retains similar (i.e., slightly decreases by 0.1\%) given a small 3\% poison rate---this 3\% poison rate was also used in~\cite{tang2021demon} to carry out SCSB attack under VCB in the same data outsourcing scenario.

\subsection{Model Outsourcing}\label{sec:modeloutsource}
In the model outsourcing scenario, the attacker can follow the normal training process as data outsourcing in \autoref{sec:DataOutsource} without \textit{interfering with the training procedure}. Note that this is not a preferable approach for the attacker as the attacker can indeed manipulate the training procedure. Therefore, we can leverage the loss regularization to improve HCB attack performance. The first regularization is expressed as: 

\begin{equation}\label{eq:loss1}
L_{1}=\sum_{(x_{i},y_{i})\in\mathcal{D}}^{}f_{\rm bd}(x_{i},y_{i};\theta)+ \sum_{(x_{i}^{'},y_{i})\in\mathcal{D^{\rm ne}}}^{}\alpha\cdot
f_{\rm bd}(\textsf{T}(x_i),y_{i};\theta).
\end{equation}

The first term means that, for non-trigger-carrying samples, the backdoored model should behave normally for them---classifying them into their ground-truth classes. This term is to ensure that i) a CDA of the backdoored model is comparable to that of the clean model and ii) a low FPR$_{\rm ES}$. The second term means that for non-effective samples \textit{with} triggers, the backdoored model should still behave normally to retain a low FPR$_{\rm NES}$. The $\alpha$ is a regularization factor, which can be set empirically for optimization. We empirically set it as 0.3 (i.e., after a few trials of factor tuning) in the following experiments unless otherwise specified---this is practical in model outsourcing as the attacker controls the training. Note merely using \autoref{eq:loss1} is insufficient, as it has not optimized the ASR. To do so, the following loss is taken:

\begin{equation}\label{eq:lossbd}
L_{\rm bd}=\sum_{(\textsf{T}(x_{i}),y_{t})\in\mathcal{D}^{e}}f_{\rm bd}(\textsf{T}(x_{i}),y_{t};\theta).
\end{equation}

The $L_{\rm bd}$ enforces that for effective samples stamped with triggers, the backdoored model should always classify them to the attacker's pre-specified target class $y_t$ to achieve a high attack success rate.

\begin{figure}[h]
	\centering
	\begin{minipage}{1\linewidth}
		\centering
		\subfigure{
			\includegraphics[width=0.47\linewidth]{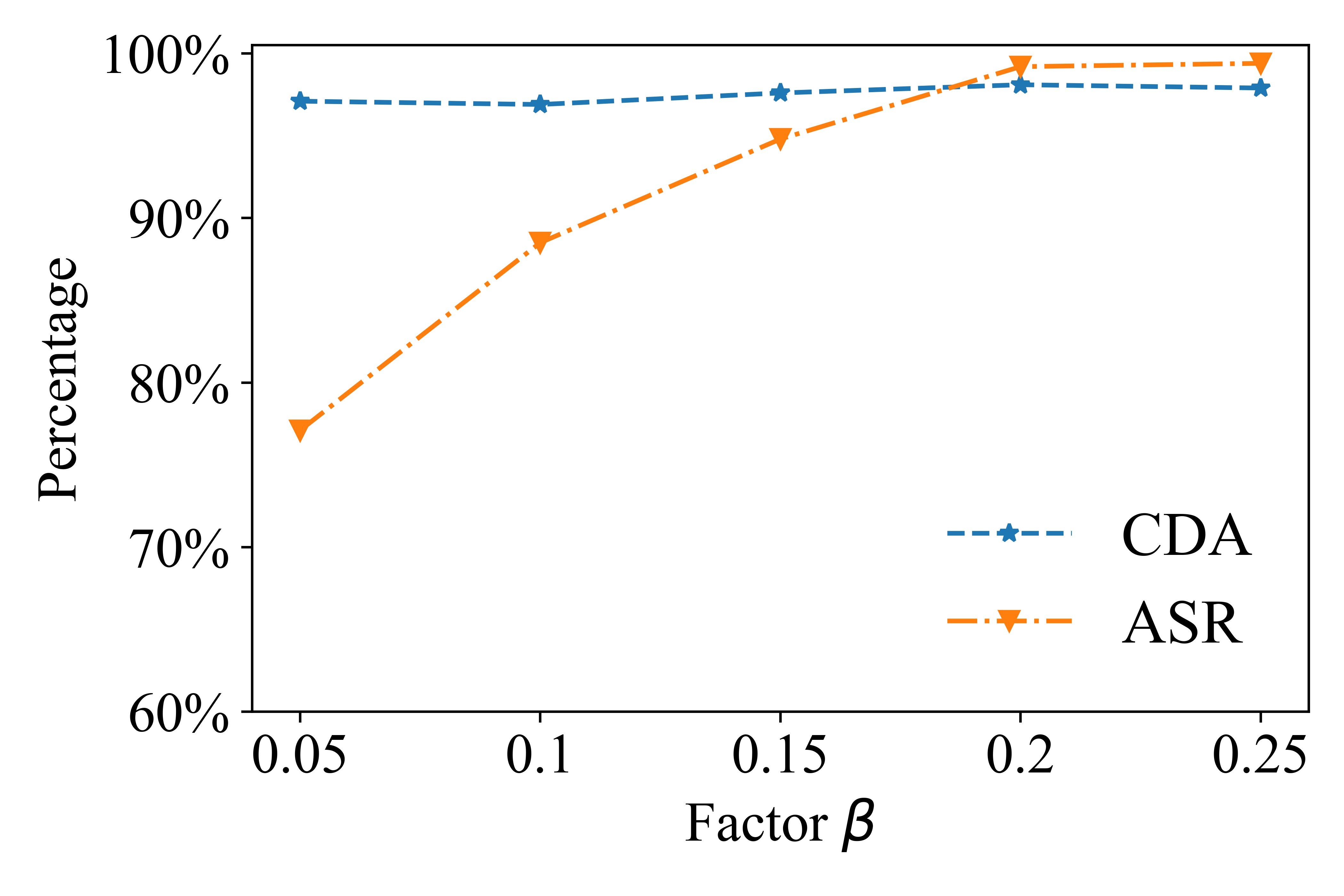}}				
		\subfigure{
			\includegraphics[width=0.47\linewidth]{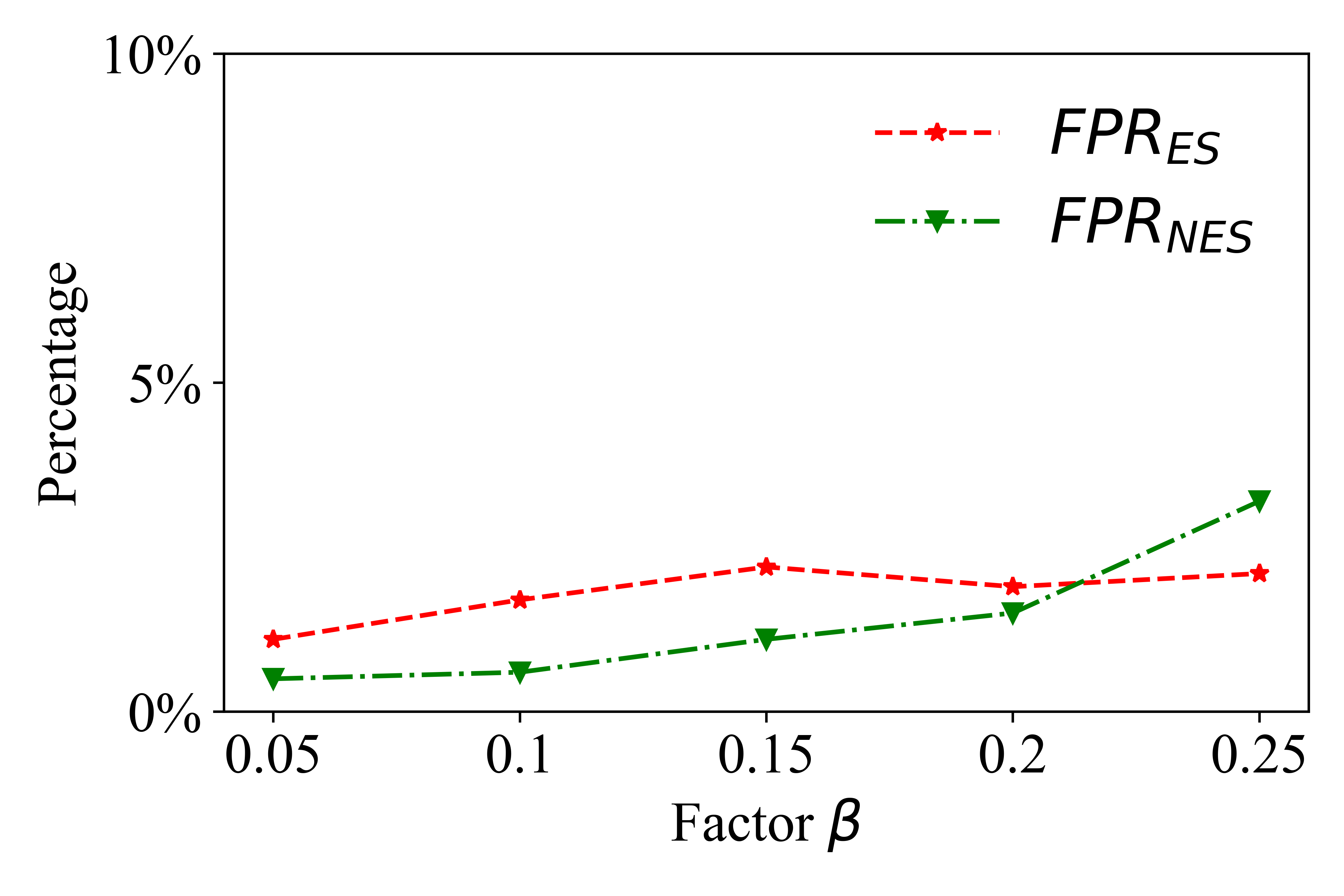}}
		\caption{The impact of factor $\beta$ on HCB performance in model outsourcing. GTSRB + rain as the innocuous feature.}
	\label{fig:beta}
	\end{minipage}
\end{figure}

The final loss constraint is then expressed as:
\begin{equation}\label{eq:total}
L=L_{1}+\beta\cdot L_{\rm bd},
\end{equation}
where $\beta$ is a regularization factor, which is empirically set as 0.2 in the following experiments unless otherwise stated. Generally, when $\beta$ increases, the backdoor behavior is gradually enhanced, but the backdoored model may over-learn the mapping relationship between effective samples stamped with triggers and $y_{t}$. In other words, the backdoor behavior might be activated by either the trigger or the innocuous features, making not only ASR, but also FPR$_{\rm ES}$ and FPR$_{\rm NES}$ increase upon increasing $\beta$ as shown in \autoref{fig:beta}---note the FPR$_{\rm ES}$ and FPR$_{\rm NES}$ are desired to be low. Nonetheless, the final loss constraint emphasizes the main task with negligible effect on the CDA.

\noindent{\bf Two-Step Attack.} Despite its feasibility of directly backdooring the model according
to~\autoref{eq:total}, dubbed as a one-step attack to ease description, we noted that the attack performance is unstable, detailed in \autoref{app:onestep}. To achieve stable attack performance, we take a two-step attack. More specifically, at the first step, we train the model according to \autoref{eq:loss1} to ensure a CDA similar to the clean model counterpart. In the second step, we fine-tune the model gained in the first step to achieve a high ASR according
to \autoref{eq:total}. With this two-step attack, the HCB attack can achieve high and stable attacking performance.

The results are detailed in Table~\ref{tab:HCBAttack}. Clearly, all four metrics' performance is improved at the same time compared to the data outsourcing scenario. Specifically, the ASR is always close to 100\%. For example, the ASR of CelebA (Smile) is improved from 92.6\% to up to 99.2\%, which increases by 6.6\%. Most importantly, both the FPR$_{\rm ES}$ and FPR$_{\rm NES}$ are greatly reduced, no more than 3\% in any case. As a comparison, the highest FPR$_{\rm ES}$ and FPR$_{\rm NES}$ are up to 7.5\% (CelebA + eye-glasses) and 9.6\% (CelebA + mouth-open), respectively, in the data outsourcing case (or without applying attack regularization in the model outsourcing case). 
In summary, the HCB attack achieves nearly ideal performance in model outsourcing with our two-step model manipulation attack. That is, the backdoored model has a CDA comparable to its clean model counterpart, an ASR close to 100\%, a close to 0\% of FPR$_{\rm ES}$ and FPR$_{\rm NES}$.

\section{Against Countermeasures}\label{sec:againstCountermeasure}
Evaluated countermeasures can be generally divided into two categories: data-level defenses and model-level defenses~\cite{gao2020backdoor}. Data-level defenses can be further performed mainly online (i.e., STRIP~\cite{gao2019strip}, Februus~\cite{doan2020februus}) or offline (i.e., SCAn~\cite{tang2021demon}, Beatrix~\cite{ma2022beatrix}), while the model-level defenses (i.e., Fine-Pruning~\cite{liu2018fine}, \textcolor{myblue}{NAD~\cite{li2020neural},} Neural Cleanse~\cite{wang2019neural}, ABS~\cite{liu2019abs}, MOTH~\cite{tao2022model}, MNTD~\cite{xu2021detecting}, MM-BD~\cite{wang2023mm}) are often taken during offline\footnote{Strictly, the online methods can always be used during offline whenever accessing the full training set is available. The Beatrix has been shown to be effective for online detection as well.}. 
Among these evaluated defenses, we note that the Februus, SCAn, Beatrix and MM-BD~\cite{wang2023mm}, by their design, are supposed to be effective against SCSB attacks. The MNTD and MM-BD are generic to the backdoor and trigger types by their design. 
Results of \textit{some defenses} are summarized\footnote{\textcolor{myblue}{Different defenses adopt varying metrics, which are challenging to be unified.}} in Table~\ref{tab:Defense}, while the results of the rest are elaborated within descriptions per defense. Due to page limitation, each defense methodology is detailed in \autoref{app:method}.

\begin{figure*}[h]
	\centering
	\setlength{\abovecaptionskip}{-0.1cm}
	\setlength{\belowcaptionskip}{-0.2cm} 
	\begin{minipage}[l]{1\textwidth}
		\centering
		\subfigure{
			\includegraphics[width=0.32\linewidth]{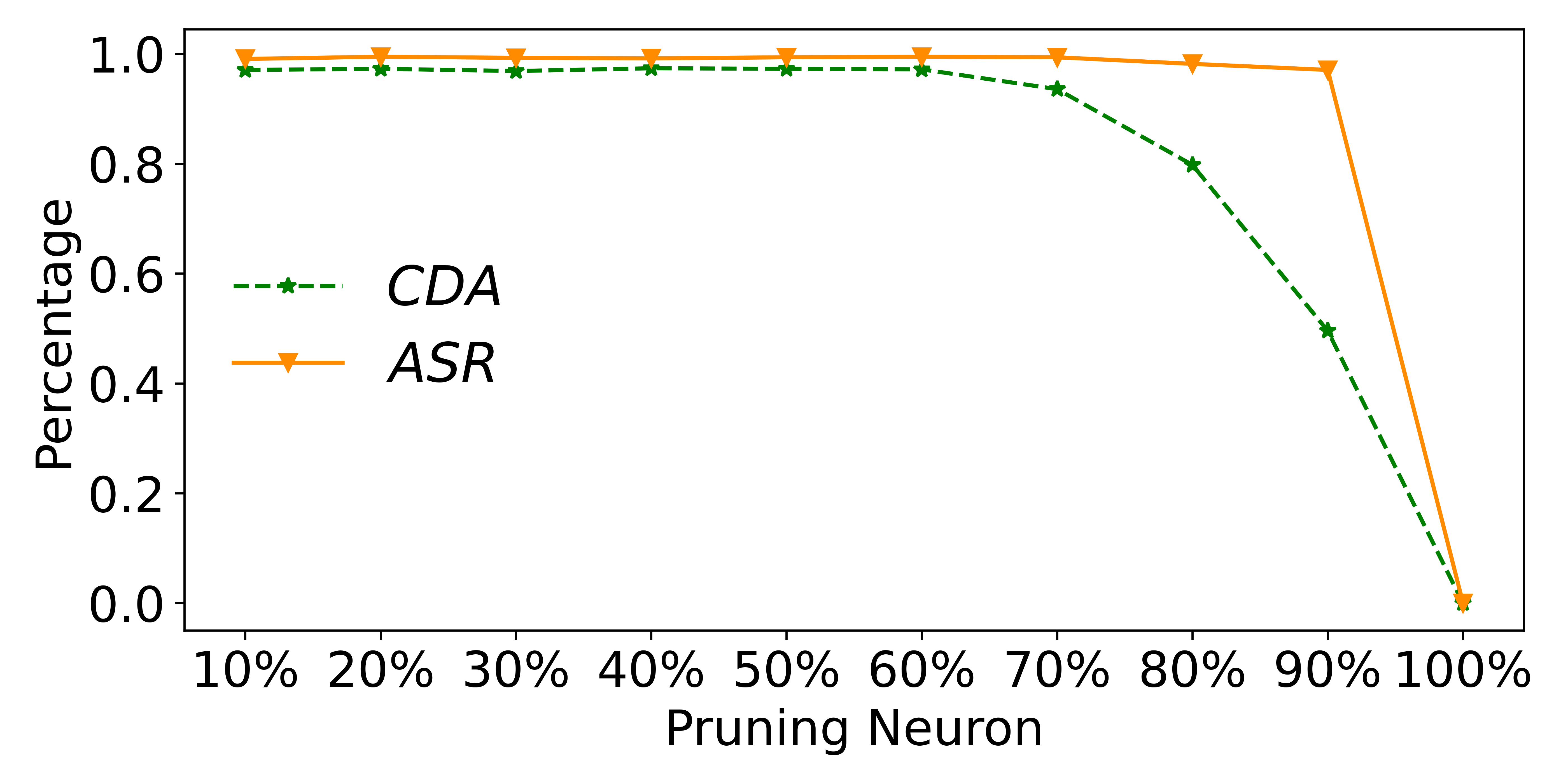}}				
		\subfigure{
			\includegraphics[width=0.32\linewidth]{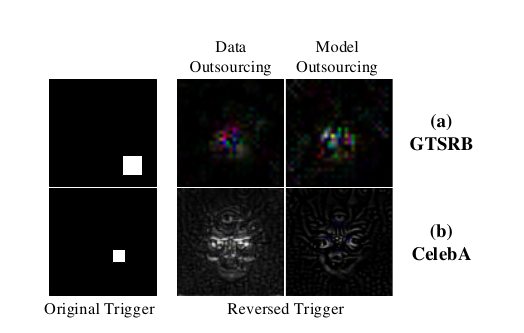}}
		\subfigure{
			\includegraphics[width=0.30\linewidth]{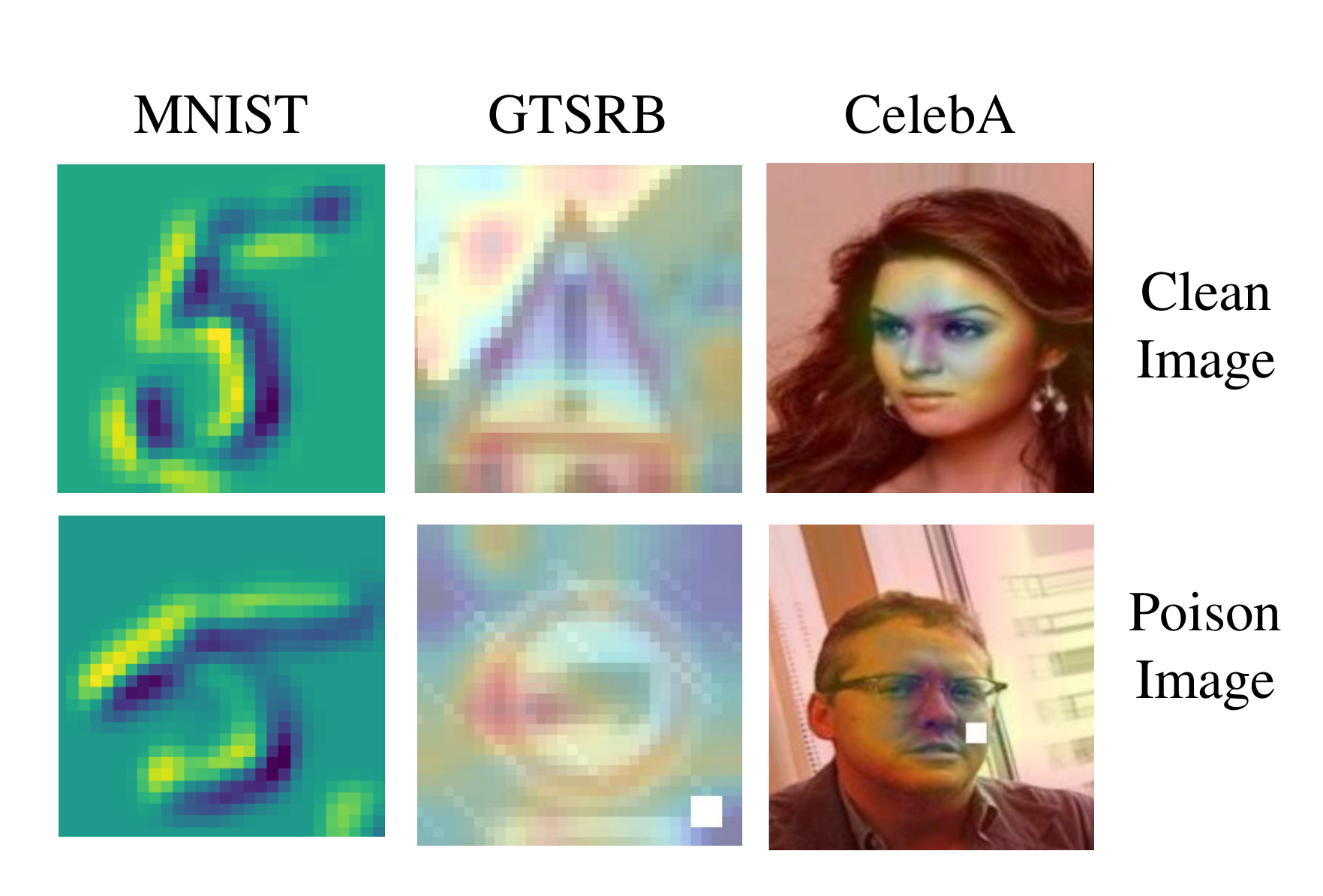}}
		\caption{(Left) The fine-pruning against the HCB attack under model outsourcing. GTSRB + rain as the innocuous feature. (Middle) The triggers reverse-engineered by Neural Cleanse does not resemble original triggers shown in the left neither from the pattern nor the position. (Right) The heatmap of clean and trigger-carrying images when GradCAM is applied.}
	\label{fig:modeldefense}
	\end{minipage}
\end{figure*}

\subsection{Model-level Defenses}\label{sec:modeldefense}

\noindent{\bf Fine-Pruning.} 
The methodology description is detailed in \autoref{app:fineprune}.
\autoref{fig:modeldefense} (Left) depicts CDA and ASR performance as a relationship of the proportion of pruned neurons in the fully connected layer. The GTSRB dataset is used and rain is the innocuous feature, we omitted results of other datasets and innocuous feature combinations because of similar tendencies. We can see that fine-pruning is unable to mitigate the HCB attack effect. Only when an almost 90\% pruning rate is applied does the ASR start dropping. However, the CDA sees unbearable degradation already. In fact, the CDA degrades first before ASR is suppressed by increasing the pruning rate.

\textcolor{myblue}{The results above were obtained when both effective and non-effective samples were employed for fine-pruning the model. However, given the rarity of effective samples and their potential exclusion from the small fine-pruning dataset, we conducted fine-pruning without effective samples. This approach is anticipated to facilitate the forgetting of innocuous features to some extent, thereby aiding in mitigating the backdoor effect. Subsequently, we conducted experiments in this setting.
Notably, the fine-pruned model exhibits a similar variation trend to that depicted in \autoref{fig:modeldefense} (Left). Notably, the ASR decreases to 86.1\%, and the CDA drops to 79.2\% when pruning 95\% of neurons in the fully connected layer. In summary, fine-pruning, irrespective of the inclusion of effective samples, proves ineffective against HCB attacks unless accompanied by significant sacrifices in model utility.
}

\noindent{\bf NAD.} \textcolor{myblue}{The methodology description is detailed in \autoref{app:nad}. We evaluated NAD\footnote{\textcolor{myblue}{We reproduced the NAD from the released source code at \href{https://github.com/bboylyg/NAD}{https://github.com/bboylyg/NAD}.}} on GTSRB with the rain as an innocuous feature under both the data outsourcing and model outsourcing scenarios. Per attack scenario per task, we trained five HCB backdoored models, where both the trigger location and attacker-target label vary per combination. Then, we leveraged the NAD to erase backdoors from these impaired models. Specifically, we assumed a powerful defender could access full clean samples instead of 5\% clean samples and follow the remaining settings (e.g., epochs, batch size and the selected layers) in the NAD's source code. For each attack scenario per task, we averaged the CDA and ASR of five repaired models.}

\textcolor{myblue}{For the data/model outsourcing scenario, the CDA and ASR are 98.12\%/98.27\% and 92.76\%/91.44\%, respectively. It indicates that NAD can slightly restrain the HCB effect under a strict defense assumption. Note that the original NAD that applies 5\% clean samples to perform the distillation operation fails to defeat the HCB attack, and the remaining ASRs are over 99\%. It is noted that the NAD costs overhead about half of that training-from-scratch, which is impractical for some resource-limited defenders. The NAD utilizes a distillation process to deliver the intermediate attention layer of the teacher network to the student network. It helps the student network to learn feature representations of clean samples, thus erasing wrong associations introduced by trigger samples. However, innocuous features entangled with clean samples' features, resulting in most of the neurons being responsible for both the main task and the backdoor task. When distilling knowledge about clean samples from the teacher network, the student network can still learn the HCB attack effect. Therefore, the NAD falls short in countering the HCB attack.}

\noindent{\bf Neural Cleanse.} The methodology description is detailed in \autoref{app:neural}.
We evaluated Neural Cleanse\footnote{We reproduced the Neural Cleanse with released source code at \href{https://github.com/bolunwang/backdoor}{https://github.com/bolunwang/backdoor}.} on GTSRB+rain and CelebA+eye-glasses tasks backdoored models under both the data outsourcing and model outsourcing scenarios.
Per attack scenario per task, we trained five HCB backdoored models, where both the trigger location and attacker-targeted label vary per combination. Then, we repeated the Neural Cleanse 20 times for each of the backdoored models. More precisely, we randomly selected 20\% of testing data containing some effective samples and applied 50 optimized epochs to reconstruct candidate triggers via the reverse-engineering process. For each attack scenario per task, we recorded the undetected rate (i.e., a backdoored model is determined as a clean model) and the average anomaly index; the results are shown in Table~\ref{tab:NC}. 

It can be seen that Neural Cleanse completely fails (the undetected rate is 100\% in our experiments) to detect HCB attack as it neither correctly reverse-engineers a trigger similar to the real one (i.e., the simplest small white-square we intentionally used) nor identifies the targeted class. Because the HCB attack trivially breaches the existence of the short-cut assumption given all non-targeted classes that are required by the Neural Cleanse, \textit{not all samples with the trigger will exhibit a backdoor effect in the HCB attack}. Therefore, the Neural Cleanse fails to reconstruct the trigger that is even the simplest---small and static white-square trigger we intentionally chose, see exemplified reconstructed (incorrect) triggers in \autoref{fig:modeldefense} (Middle).

\begin{table}[]
\caption{HCB evasiveness against Neural Cleanse.}
\label{tab:NC}
\centering
\resizebox{0.5 \textwidth}{!}
{
\begin{tabular}{c|c|cc}
\hline
Task   & Metric                                                                                       & Data Outsourcing & Model Outsourcing \\ \hline
GTSRB  & \multirow{2}{*}{\begin{tabular}[c]{@{}c@{}}Undetected Rate \\ \& Anomaly Index\end{tabular}} & 100\% \& 1.55         & 100\% \& 1.62          \\
CelebA &                                                                                              & 100\% \& 1.72            & 100\% \& 1.67             \\ \hline
\end{tabular}
}
\end{table}

\noindent{\bf MOTH.} The methodology description is detailed in \autoref{app:moth}.
With MOTH\footnote{We reproduced the MOTH with released source code at \href{https://github.com/Gwinhen/MOTH}{https://github.com/Gwinhen/MOTH}.}, we evaluated the backdoored models on MNIST+black-paper, GTSRB+rain and CelebA+eye-glasses tasks under both the data outsourcing and model outsourcing scenarios. Per attack scenario per task, we trained five HCB-impaired models. Then, we used the MOTH to harden these impaired models. We randomly selected 5\% training data 
to reconstruct trigger candidates for each of the selected class pairs, which technique is followed by~\cite{tao2022model}. For each attack scenario per task, we report the average ASR of five hardened/repaired models as the remaining ASR, the results are shown in Table~\ref{tab:Defense}. 
Though MOTH costs substantial computing resources to enlarge the class distance, it is ineffective to HCB-infected models. More precisely, in the MNIST+black-paper task, the ASRs after repairing in the data and model outsourcing scenarios are 94.7\% and 97.3\%, respectively, which means that MOTH completely fails. 

For the rest evaluated tasks, although MOTH could reduce the backdoor effect to some extent, there is a notable trade-off between the degree of ASR mitigation and the deterioration of CDA.
For instance, it can erase the HCB attack effect on the GTSRB under model outsourcing, exhibiting a desirable ASR of 0\%. However, the repaired model has an average CDA of only 47.4\% (i.e., the clean model CDA is 98.2\%, see Table~\ref{tab:HCBAttack}), rendering the utility useless.
We analyze the potential reason as below. Considering the model outsourcing, we delicately design the loss function for the sake of higher ASR and lower FPR$_{\rm NES}$, which differs from the normal backdoor training (e.g., under data outsourcing without training process manipulation). The MOTH seems not optimized against the case when the backdoor insertion is delicately done through regulations. Therefore, the MOTH severely undermines both the main and backdoor tasks. 

As for the CelebA+eye-glasses task, the MOTH demonstrates a similar tendency to the GTSRB. 
Despite the ASR drops to 50.9\% and 48.6\% for data outsourcing and model outsourcing, respectively. The CDA of the repaired model has been reduced by notably 11.1\% (i.e., from 95.9\% to 84.8\%) and 12.1\% (i.e., from 97.3\% to 85.2\%), respectively.

There are two major reasons for MOTH failure. Firstly, the reverse-engineered trigger is class-dependent---it relies on the technique of Neural Cleanse~\cite{wang2019neural} for trigger synthesis. That is, samples from one class will be misclassified into the other class once the trigger is attached. However, this is not true given the HCB due to the backdoor being also entangled with the innocuous feature of a class. Secondly, the MOTH is mainly assuming that the trigger is static. Following the above reason, the innocuous feature can be generally considered as a partial component of the trigger for \textit{effective samples} to activate the HCB backdoor. Therefore, the HCB can already fail to synthesize a meaningful trigger/perturbation that is important for the following model repairing/hardening process.

\begin{table}[]
\caption{HCB evasiveness against backdoor countermeasures.}
\label{tab:Defense}
\centering
\begin{threeparttable}
\resizebox{0.50 \textwidth}{!}
{
\begin{tabular}{c|c|ccc|ccc}
\hline
\multirow{2}{*}{Defense} & \multirow{2}{*}{Metric} & \multicolumn{3}{c|}{Data Outsourcing} & \multicolumn{3}{c}{Model Outsourcing}    \\ &
                         & MNIST       & GTSRB     & CelebA     & MNIST            & GTSRB            & CelebA$^1$  \\ \hline
{\begin{tabular}{@{}c@{}} Februus~\cite{doan2020februus} \\ (ACSAC 20')    \end{tabular}} & Remained ASR               & 99.8\%           & 90.5\%         & 85.4\%          & 99.7\%               & 93.7\%               & 90.8\%  \\ \hline

{\begin{tabular}{@{}c@{}} SCAn~\cite{tang2021demon} \\ (Usenix 21')    \end{tabular}} & Undetected Rate                 & 90.3\%           & 62.2\%         & 71.8\%          &N/A & N/A & N/A  \\\hline

{\begin{tabular}{@{}c@{}} MNTD~\cite{xu2021detecting} \\ (Oakland 21')    \end{tabular}} &  {\begin{tabular}{@{}c@{}} False Rej. Rate \\ False Accept. Rate   \end{tabular}}                     & {\begin{tabular}{@{}c@{}} 53.8\% \\ 52.1\%   \end{tabular}}           & {\begin{tabular}{@{}c@{}} 57.1\% \\ 53.8\%   \end{tabular}}        & {\begin{tabular}{@{}c@{}} 62.3\% \\ 52.8\%   \end{tabular}}         & {\begin{tabular}{@{}c@{}} 50.4\% \\ 47.8\%   \end{tabular}}               & {\begin{tabular}{@{}c@{}} 51.5\% \\ 51.8\%   \end{tabular}}               & {\begin{tabular}{@{}c@{}} 56.7\% \\ 57.9\%   \end{tabular}}  \\\hline

{\begin{tabular}{@{}c@{}} MOTH~\cite{tao2022model} \\ (Oakland 22')     \end{tabular}}         & {\begin{tabular}{@{}c@{}} Remained ASR \\ Repaired CDA   \end{tabular}}        & {\begin{tabular}{@{}c@{}} 94.7\% \\ 99.1\%   \end{tabular}}     & {\begin{tabular}{@{}c@{}} 91.6\% \\ 90.2\%   \end{tabular}}         & {\begin{tabular}{@{}c@{}} 50.9\% \\ 84.8\%   \end{tabular}}          & {\begin{tabular}{@{}c@{}} 97.3\% \\ 99.4\%   \end{tabular}}            & {\begin{tabular}{@{}c@{}} 0\% \\ 47.4\%   \end{tabular}}               & {\begin{tabular}{@{}c@{}} 48.6\% \\ 85.2\%   \end{tabular}}        \\\hline


 {\begin{tabular}{@{}c@{}} Beatrix~\cite{ma2022beatrix} \\ (NDSS'23)    \end{tabular}}& Undetected Rate & 100\%           & 82\%               & 0\%$^{2}$ & 87.3\%      & 71.4\%         & 93.8\%    \\ \hline
 
\end{tabular}
}
\begin{tablenotes}
        \footnotesize
        \item[1] We select eyeglasses as the innocuous feature for evaluations.
        \item[2] Note this is accompanied by undesirable severe false negatives.
      \end{tablenotes}
    \end{threeparttable}
\end{table}

\noindent{\bf MNTD.} The methodology description is detailed in \autoref{app:mntd}.
We evaluated the MNTD on the same HCB tasks as the MOTH evaluation. There are $2\times M$ shadow models trained in the offline phase. Note $M$ is 1024, 512, and 256 for  MNIST, GTSRB, and CelebA tasks, respectively. Concretely, we obtained $M$ benign models through training on the clean dataset with different model hyper-parameter initialization. For $M$ backdoored models, we followed the jumbo learning in MNTD~\cite{xu2021detecting} to generate 
backdoored models. Following the MNTD's setup, we applied modification attacks (i.e., trigger patch replace the original image content) and blend attacks (i.e., trigger occupies full image size with some transparency) when poisoning samples to insert a backdoor. 
In the detection phase, we used the median of all the model-under-test (MUT) scores (essentially logits value) returned by the binary meta-classifier as the threshold to differentiate the benign model and the backdoor model. 
Per attack scenario per task, we used 128 benign models and 128 HCB-infected models with convergence as the testing MUT dataset and ran the MNTD five times to obtain the average FAR and FRR.

The results of MNTD\footnote{We reproduced the MNTD with released source code at \href{https://github.com/AI-secure/Meta-Nerual-Trojan-Detection}{https://github.com/AI-secure/Meta-Nerual-Trojan-Detection}.} are shown in Table~\ref{tab:Defense}, we can see that the MNTD fails to distinguish backdoored models from clean models, its detection performance is only slightly better than guessing. Note the MNTD has both close to 50\% FAR (i.e., regarding a clean model as backdoored) and FRR (i.e., regarding a backdoored model as clean). The failure is mainly rooted in the MNTD's reliance on the logits distribution which is an \textit{unreliable} component. Because it is highly sensitive to the hyperparameters (i.e., number of epochs, batch size) used to train the MUT. Once these hyperparameters are varying (always the case in practice) from those used to train the shadow models, the meta-classifier will fail. The non-robustness of the MNTD has been recently noticed and evaluated by other researchers~\cite{qiu2023towards}.

\begin{figure}[t]
	\centering
	\includegraphics[trim=0 0 0 0,clip,width=0.5\textwidth]{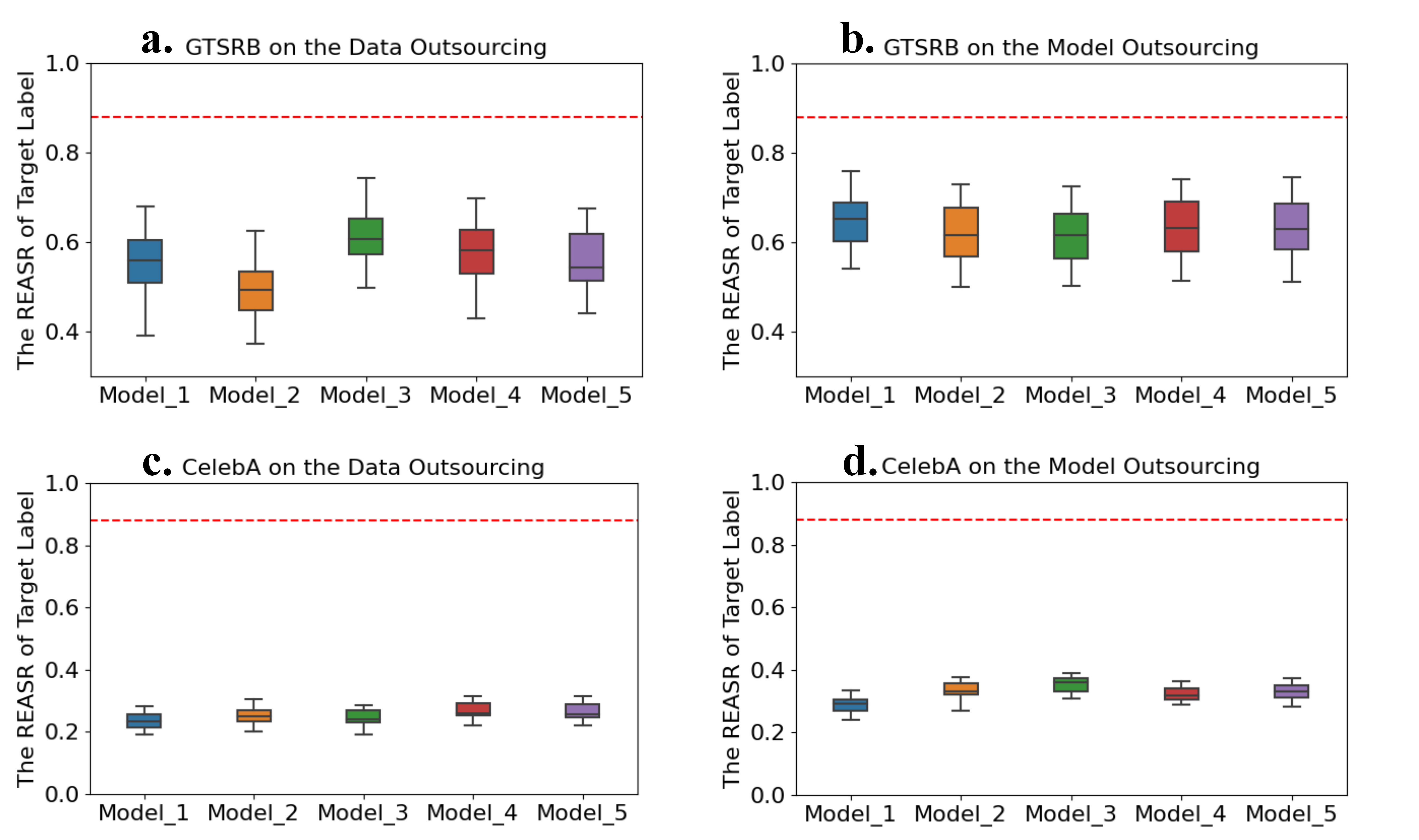}
	\vspace{-1mm}
	\caption{The REASR of ABS against the HCB-infected target label. Per attack scenario per task, five infected models are tested. Each model is repeatedly evaluated 20 times. The red dotted line is the threshold of 0.88.}
	\label{fig:abs}
	\vspace{-4mm}
\end{figure}

\noindent{\bf ABS.} The methodology description is in \autoref{app:abs}.
It has been acknowledged~\cite{liu2019abs} that one of its limitations is the assumption that the backdoor behavior is dominated by a single or only a few neurons. This assumption tends to be often not met in practice~\cite{qiu2023towards,tang2021demon}.
We evaluated ABS\footnote{We reproduced the ABS with released source code at \href{https://github.com/naiyeleo/ABS}{https://github.com/naiyeleo/ABS}.} on the same HCB attacked tasks as the Neural Cleanse. For GTSRB/CelebA, we picked up 10/5 samples from each class for ABS reverse-engineer process. Note ABS uses the reverse-engineer attack success rate (REASR) as a backdoor indicator. The threshold of 0.88 was used~\cite{liu2019abs}, where the model is backdoored if the REASR is higher than the threshold. Per the attack scenario per task, we used the ABS to detect five different HCB infected models and repeated the ABS detection 20 times per model. The detection results are depicted in \autoref{fig:abs}. We can see that ABS fails against HCB attack under any attack scenario, which is not unexpected. Because the ABS is designed with an assumption on its mainly targeted source-class-agnostic attack. Only under such conditions, can a single or few neurons dominate the backdoor effect. The HCB attack trivially invalidates the ABS premise, thus breaking it.

\noindent{\bf MM-BD.} The methodology description is in \autoref{app:mmbd}. 
We evaluated MM-BD on the same GTSRB task as the Neural Cleanse. Per attack scenarios, we used MM-BD\footnote{We reproduced the MM-BD with released source code at\url{https://github.com/wanghangpsu/MM-BD}} to detect ten HCB models trained with a random target class. 
MM-BD misjudges three infected models as clean models and still leaves up to 30\% ASR after applying its mitigation.
For misjudged cases, the HCB models did not show sufficient high ASR (i.e., 92\%). Then, we obtained ten new infected HCB models by slightly upper bound their ASR to be around 90\%. Here, MM-BD cannot detect any infected HCB models. Overall, MM-BD is not robust against our HCB attacks.

\subsection{Data-level Defenses}
The data-level defenses can be performed online and offline. The former detects (i.e., STRIP) or suppresses (i.e., Februus) the backdoor behavior given an incoming input fed into the deployed (backdoored) model---requiring no access to the training dataset. In contrast, the latter (i.e., SCAn and Beatrix) assumes full training dataset access to identify and remove poisoned samples before conducting model training.

\begin{figure}[h]
	\centering
	\includegraphics[trim=0 0 0 0,clip,width=0.45\textwidth]{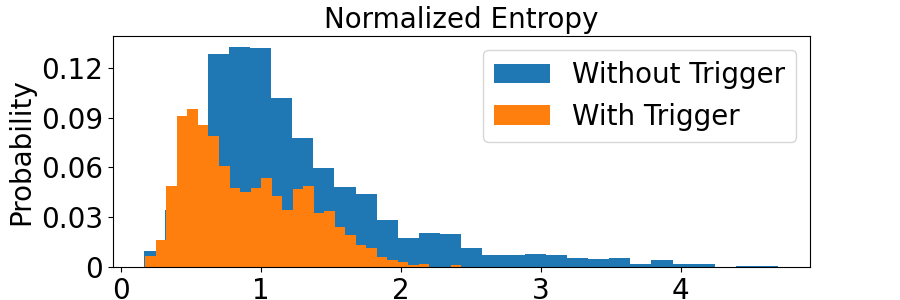}
	\vspace{-1mm}
	\caption{Entropy distributions of STRIP against the HCB attack under data outsourcing scenario. GTSRB + rain is the innocuous feature.}
	\label{fig:strip}
	\vspace{-4mm}
\end{figure}

\noindent{\bf STRIP.} The methodology description is detailed in \autoref{app:strip}.
For GTSRB/CelebA, we have used 1000/300
effective trigger-samples and 1000/400 benign samples to evaluate the STRIP\footnote{We reproduced the STRIP with released source code at \href{https://github.com/garrisongys/STRIP}{https://github.com/garrisongys/STRIP}.} detection performance. For the GTSRB data outsourcing scenario, the FAR is up to 100\%/99.5\% when the FRR is preset to be 1\%/5\%. For GTSRB model outsourcing, the FAR is 100\%/100\% given a preset FRR of 1\%/5\%. 
For CelebA Data outsourcing, the FAR is 79.3\%/67.1\% given a preset FRR of 1\%/5\%. For CelebA model outsourcing, the FAR is 76.03\%/62.9\% given a preset FRR of 1\%/5\%. 
This means the STRIP fails to detect effective trigger-carrying samples because there is a large entropy distribution overlap of effective trigger-samples and benign samples, see \autoref{fig:strip}, which results in unacceptable FAR given a reasonable preset FRR choice. The main reason is that the trigger itself is not the only dominant factor to hijack the backdoor model to make predictions, while the trigger dominant effect is the key assumption used by the STRIP. 

\noindent{\bf Februus.} The methodology description is detailed in \autoref{app:februus}.
As detailed in Table~\ref{tab:Defense}, the ASR after applying the Februus\footnote{We reproduced the Februus with released source code at \href{https://github.com/AdelaideAuto-IDLab/Februus}{https://github.com/AdelaideAuto-IDLab/Februus}.} is still high. Specifically, in the data outsourcing scenario, the ASR of MNIST, GTSRB, and CelebA are up to 99.8\%, 90.5\%, and 85\%, respectively. In the model outsourcing, the ASRs of these three tasks are even higher. Therefore, the Februus is incapable of defeating HCB attacks, even if it can defeat SCAB and SCSB attacks all under the conventional VCB. In addition, the backdoor effect in the model outsourcing scenario is harder to suppress than that in the data outsourcing scenario due to the higher controlling ability of the attacker to insert a stealthier backdoor.

Generally, the HCB attack renders the backdoor effect to be an entanglement of the innocuous feature and the trigger. Consequently, the trigger-located region on which the Februus relies is hard to be explicitly identified. We have plotted the heatmap using GradCAM for three different datasets, as in \autoref{fig:modeldefense} (Right). Three images on the left/right column are benign/trigger-carrying images. It can be seen that benign and its counterpart trigger-carrying images exhibit similar heatmaps regardless of the trigger. So Februus fails to mitigate the backdoor effect of the trigger-carrying images.

\noindent{\bf SCAn.} The methodology description is detailed in \autoref{app:scan}. Note that SCAn requires one to discern a certain number of (i.e., 50) trigger-carrying samples before reliable online detection~\cite{tang2021demon,ma2022beatrix} for the model outsourcing scenario, which is problematic as the system can already be broken~\cite{ma2022beatrix}. We thus only focus on evaluating the SCAn in the data outsourcing scenario.
Following SCAn setting~\cite{tang2021demon}, we used a small set of clean data occupying 10\% of the whole dataset. Through hypothesis testing, we obtained whether there is a possibility of multiple (i.e., two) identities in a given class (i.e., multi-distribution modeling effect). According to the source code, a score higher than the threshold of $e^{2}$ indicates the infection. Per attack scenario per task, we randomly poisoned 100 datasets with an HCB attack and applied the SCAn to detect these poisoned datasets. This procedure was repeated five times, where an average undetected rate was reported.

The results of SCAn\footnote{We reproduced the SCAn with released source code at \href{https://github.com/TDteach/Demon-in-the-Variant}{https://github.com/TDteach/Demon-in-the-Variant}.} on HCB is shown in Table~\ref{tab:Defense}. It can be seen that SCAn is ineffective in identifying HCB poison samples (in particular, trigger samples with innocuous features). There are two potential reasons. Firstly, the main limitation of SCAn is that it attempts to decompose two mean vectors (i.e., each corresponding to an identity/class) if a given class is infected. One mean vector stands for the representation of benign samples from this class, while the other mean vector indicates the poisoned samples (they are from other classes but their labels are changed to the infected class). This decomposition is thus class-dependent. The HCB breaches this assumption, thus rendering the SCAn ineffective. Secondly, the first SCAn decomposition operation is to find a universal variation from the defender held-out samples. This variation is supposed to be not related to the backdoor effect and is an important factor in performing the two mean vectors decomposition. However, the HCB has entangled the backdoor effect with this variation (i.e., mouth-open used by HCB as an innocuous feature is regarded as a universal variation by SCAn).

\noindent{\bf Beatrix.} Other detection~\cite{chen2019detecting,tang2021demon} all only utilize the first moment (mean) discrepancy of e.g., latent representation between benign and trigger samples.
Beatrix delves into higher-order information on the latent representation of benign and trigger samples to detect trigger samples in the Gramian feature space, treating the trigger sample detection as a problem of out-of-distribution (OOD) detection.

We evaluated the Beatrix\footnote{We reproduced the Beatrix with released source code at \href{https://github.com/wanlunsec/beatrix}{https://github.com/wanlunsec/beatrix}.} on the same tasks as the MOTH evaluation. For MNIST, GTSRB, and CelebA tasks under data outsourcing (detect infected class in an offline phase before model deployment) and model outsourcing scenarios (detect incoming trigger-carrying samples in an online phase after model deployment), we randomly selected 30, 30, and 5 clean samples, respectively, from each class. Because the Beatrix requires a small set of clean samples.
For infected class detection in data outsourcing scenarios, we evaluated five different HCB infected models and repeated Beatrix detection 20 times per model for all tasks. For online sample detection in model outsourcing scenarios, we evaluated 2000, 2000, and 300 benign and trigger-carrying samples for MNIST, GTSRB, and CelebA, respectively. 
The average detection rate of Beatrix on HCB is shown in Table~\ref{tab:Defense}. It can be observed that Beatrix is not effective against the HCB attacks as the undetection rate remains quite high (above 70\% in all cases). We note there is an exception for CelebA task given the infected class detection. This is caused by the unreliability of Beatrix when the number of held-out clean samples per class is low. Due to CelebA having a small number of samples per class, only 5 clean samples per class are reserved for Beatrix---the minimum functional number is 8 according to~\cite{ma2022beatrix}. 
Overall, the reason for failure is similar to that of the SCAn, as Beatrix is also a class-dependent defense. 

\section{Discussion}\label{sec:discussion}

\subsection{Model Complexity}
\begin{figure*}[t]
	\centering
	\setlength{\abovecaptionskip}{-0.1cm}
	\setlength{\belowcaptionskip}{-0.2cm} 
	\begin{minipage}[l]{1\textwidth}
		\centering
		\subfigure{
			\includegraphics[width=0.29\linewidth]{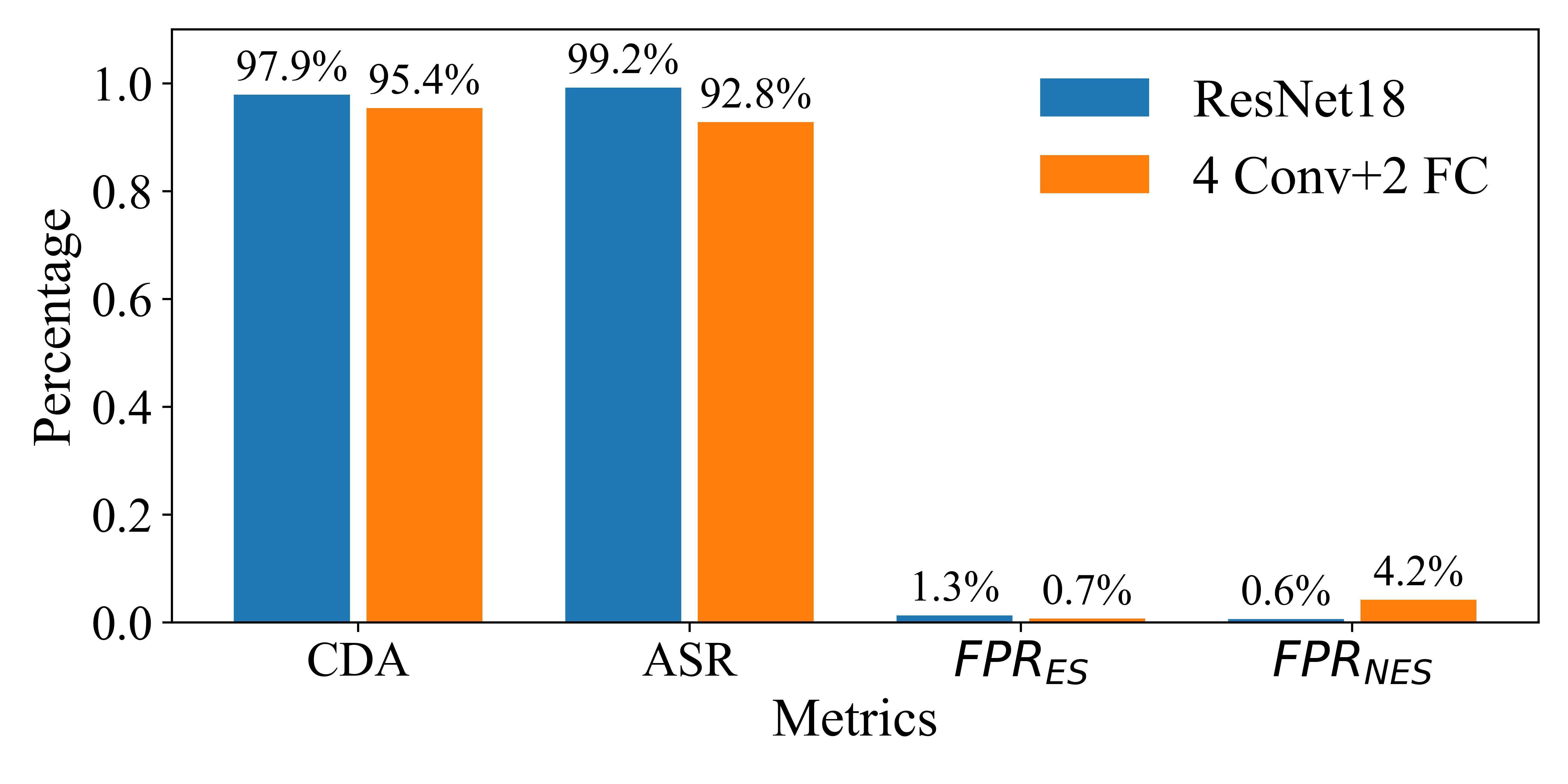}}				
		\subfigure{
			\includegraphics[width=0.38\linewidth]{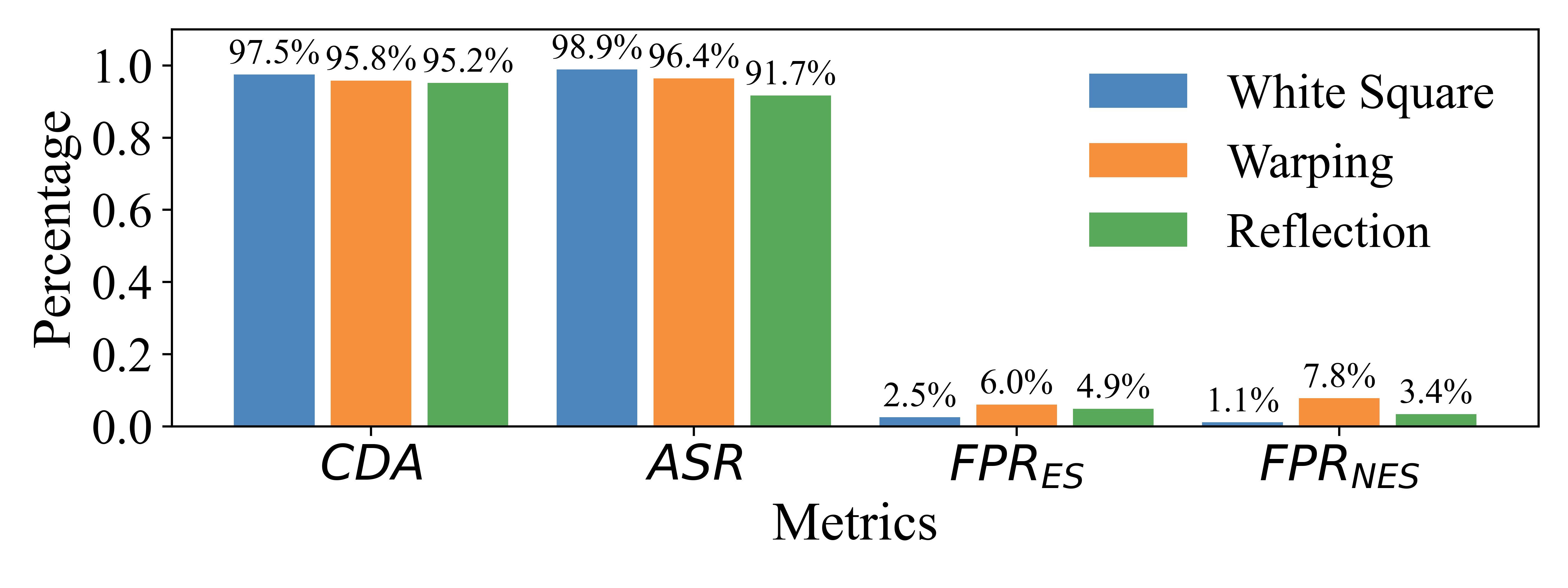}}
		\subfigure{
			\includegraphics[width=0.22\linewidth]{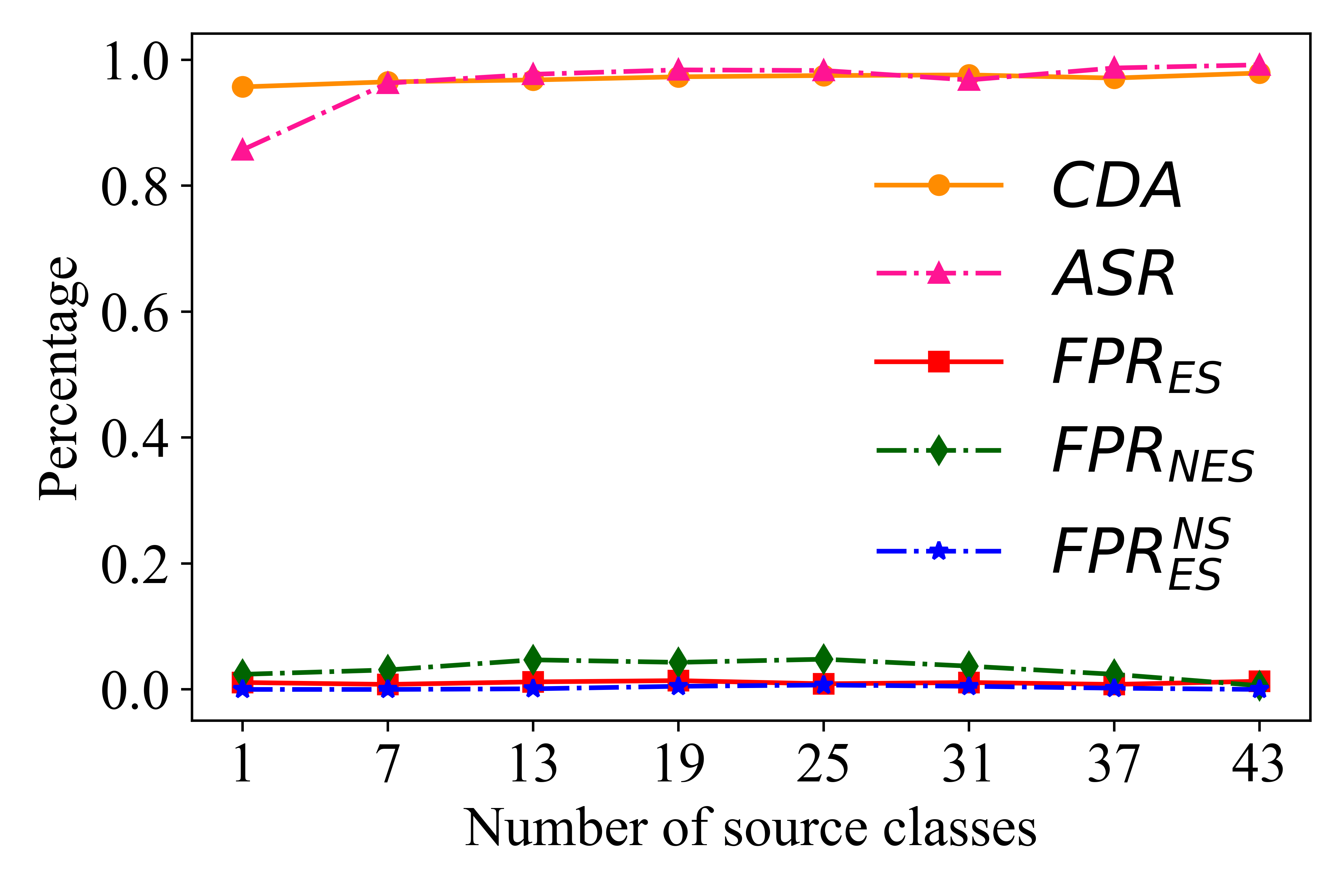}}
		\caption{(Left) The comparison of attack performance between shallow model and ResNet18 under model outsourcing. (Middle) HCB attack performance when different trigger types are incorporated. (Right) Incorporating source-specific effect into HCB attack. The GTSRB is used and rain is the innocuous feature.}
	\label{fig:discus}
	\end{minipage}
 \vspace{-4mm}
\end{figure*}

The HCB attack performance can be influenced by the model complexity. Generally, a more complex model indicates better attack performance. Because the model has improved the capability to learn both the main task and the backdoor task. In previous experiments we used ResNet18 to train GTSRB (rain as an innocuous feature), we now use a relatively shallow architecture with 4 convolutional and 2 fully connected (FC) layers. 
The results are detailed in \autoref{fig:discus} (Left). In the model outsourcing scenario, the CDA is 95.4\% (CDA of ResNet18 is 97.9\%). The ASR, FPR$_{\rm ES}$, and FPR$_{\rm NES}$ of shallow/ResNet18 model is 92.8\%/99.2\%, 0.7\%/1.3\%, 4.2\%/0.6\%. The shallow model attack performance degrades. This is reasonable because the shallow model has a relatively weak capability of learning the innocuous feature, exhibiting a decreased ASR and FPR$_{\rm ES}$. Now the trigger gains a relatively higher weight to activate the backdoor, resulting in a higher FPR$_{\rm NES}$.

\subsection{Object Detection}\label{sec:object}

\textcolor{myblue}{We note that existing backdoor attacks and defenses rarely apply to non-classification object detection tasks, particularly with natural triggers (e.g., blue T-shirt as a trigger), but we consider such a challenging task for generalization validation.}
Here, the backdoor effect is person cloaking where a person wearing a natural T-shirt (see the \textit{blue T-shirt} in \autoref{fig:object_detection} (e)) as the physical trigger disappears (i.e., no bounding box is proposed) and rain as the innocuous feature. \textit{So that the HCB-based person cloaking effect only exhibits when the person wears the trigger T-shirt and the weather is rainy.}

The training set is a mixture of the VOC 2007 training set (2,501 samples) and the VOC 2012 train-validation set (11,540 samples). In the validation phase, the VOC 2007 validation set (2,510 samples) is used. For the HCB attack, we crafted 552 dirty samples and 552 cover samples to be added to the training set. The former has rainy images, and the bounding box of people with the trigger T-shirt is removed. The latter has no-rain images, and the bounding box of the trigger person is intact. 

In the experiment, we set the $\beta$ as 0.5 in the ~\autoref{eq:total}. We use the popular object detector of Yolo-V4~\cite{bochkovskiy2020yolov4} to implement the task and set the input size as $416 \times416 \times3$. There are 20 classes of objects including a person.
To improve training efficiency, we utilize a pre-trained model that has been trained on the COCO dataset. The training is done with 200 epochs, of which the first 100 epochs are frozen training and the remaining epochs are unfrozen training. 
We leverage the common intersection over-union (IoU) in object detection for performance evaluations and set the IoU confidence as 0.5. That is, the mAP@0.5 is used. The CDA of backdoored Yolo-V4 is 85.32\% which is on par with that of a clean model counterpart. The ASR is up to 87\% (by evaluating 100 images with rain + trigger T-shirt person). The FPR$_{\rm ES}$ is 5\% (by evaluating 100 images with no-rain + trigger T-shirt person) and the FPR$_{\rm NES}$ is 4\% (100 images with rain + non-trigger T-shirt person). 
An example is shown in \autoref{fig:object_detection}.

\begin{figure}[h]
	\centering
 	\setlength{\abovecaptionskip}{-0.1cm}
	\setlength{\belowcaptionskip}{-0.2cm} 
	\begin{minipage}{1\linewidth}
		\centering
		\subfigure{
			\includegraphics[width=0.47\linewidth]{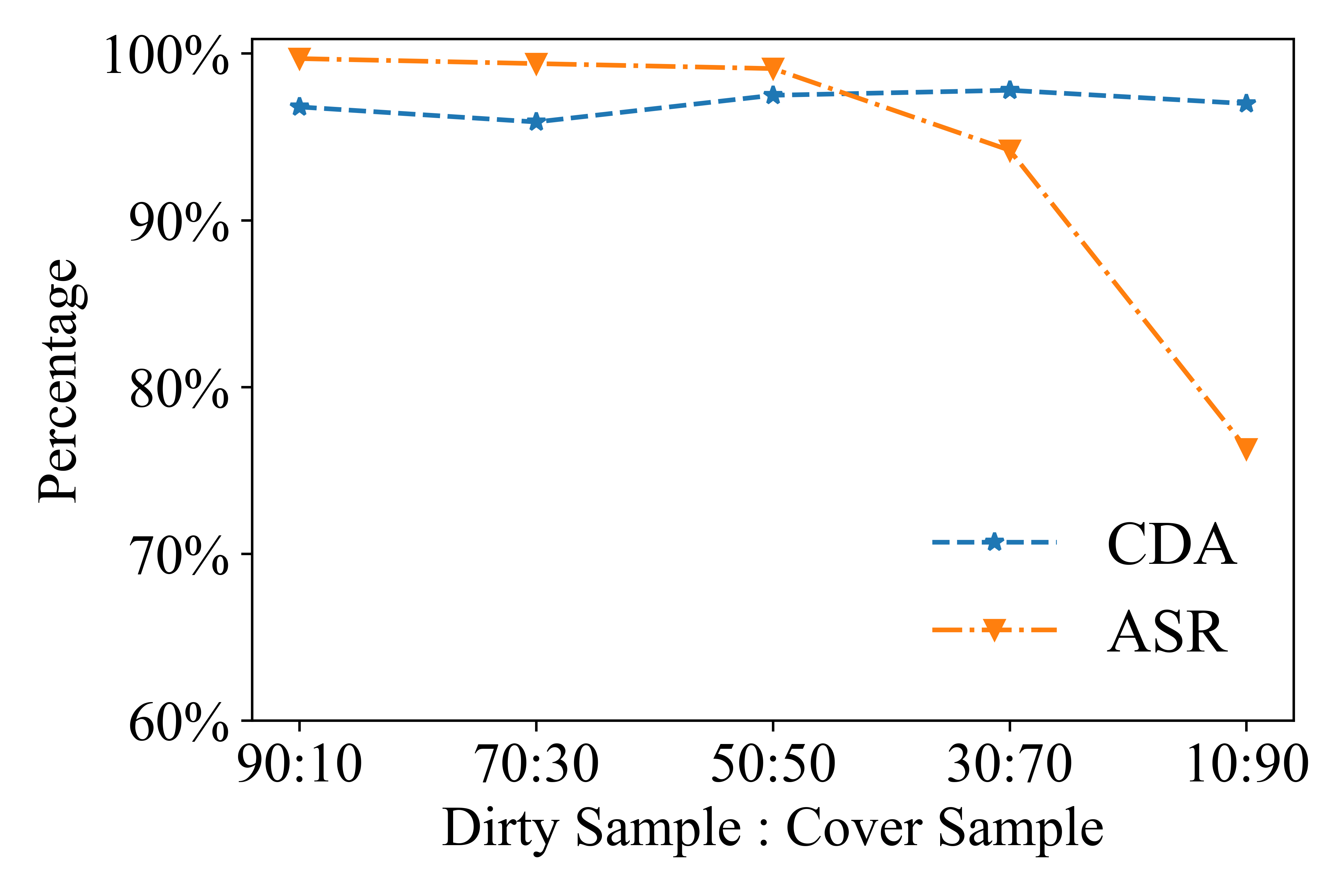}}				
		\subfigure{
			\includegraphics[width=0.47\linewidth]{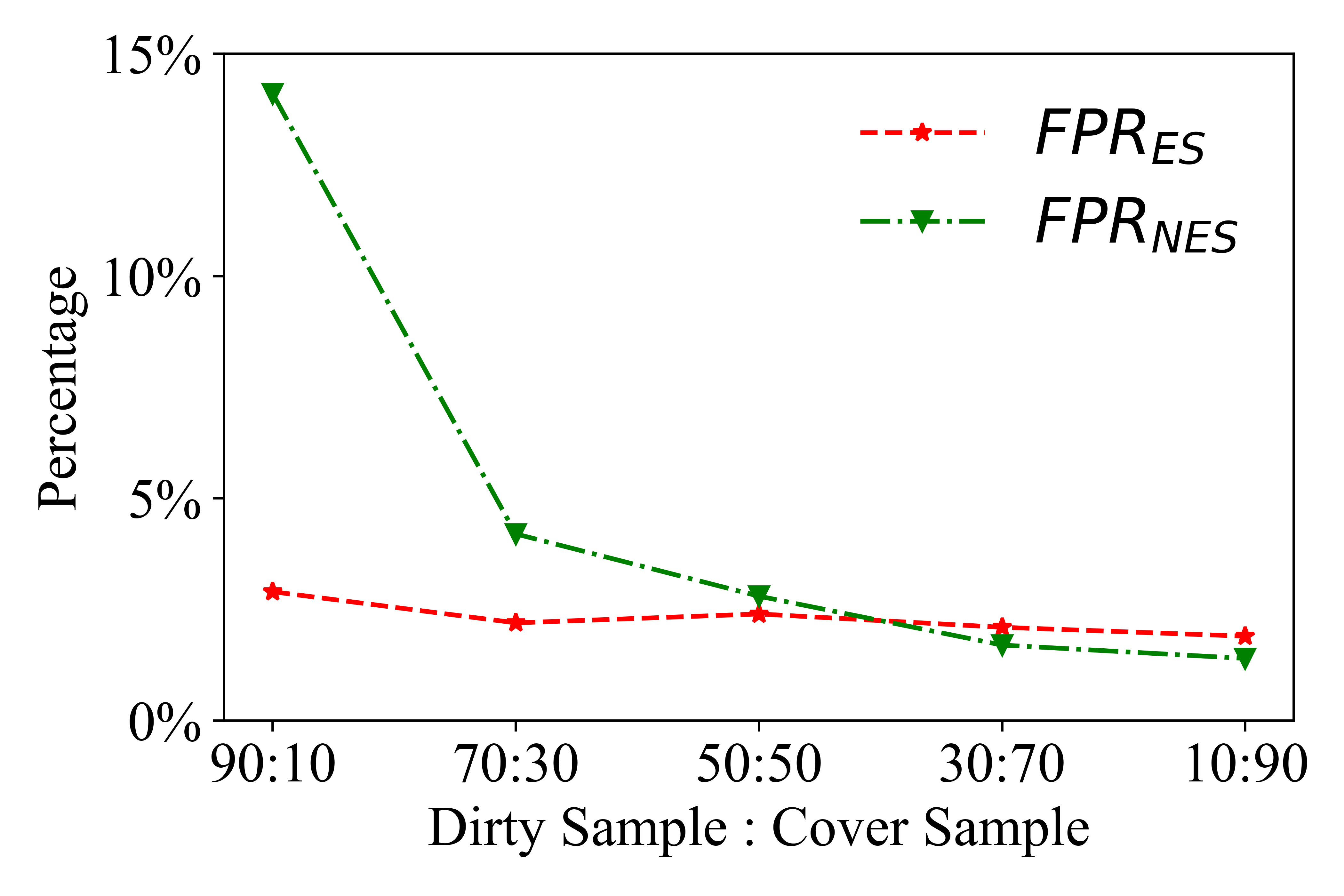}}
		\caption{HCB attack performance as a function of the ratio between dirty and cover samples. GTSRB + rain is the innocuous feature.}
	\label{fig:cover_ratio}
	\end{minipage}
\end{figure}

\subsection{\textcolor{myblue}{Ratio between Dirty and Cover Samples}}

\textcolor{myblue}{In all previous experiments, a balanced (50:50) ratio between dirty and cover samples was maintained. The former is crafted using effective samples, while the latter is derived from non-effective samples. Despite the usual scenario where the number of effective samples is significantly lower than that of non-effective samples, achieving this balanced ratio within the poisoned samples is relatively straightforward. This is primarily because the overall poison rate is very small, typically just a few percent. It's important to note that both dirty and cover samples contribute to this low poison rate.}

\textcolor{myblue}{We further investigate the performance of HCB attacks under varying ratios between dirty and cover samples, focusing specifically on the data outsourcing scenario since in the model outsourcing scenario, the adversary does not need to expose its poisoned samples. We consider the GTSRB task with rain to be an innocuous feature. The poison rate is fixed at 8\%. The ratios of dirty to cover samples vary from 90:10, 70:30, 50:50, 30:70 to 10:90. The detailed results are depicted in Figure \ref{fig:cover_ratio}.
It is observed that the ratio primarily affects the ASR and FPR$_{\text{NES}}$ performance of HCB, while the CDA and FPR$_{\text{ES}}$ are negligibly influenced. As the ratio of cover samples increases, the infected model exhibits a decreasing trend in ASR and FPR$_{\text{NES}}$. This is because an abundance of cover samples hinders the learning of backdoor patterns, making it difficult to implant an effective backdoor. Conversely, decreasing the ratio of cover samples leads to an increasing trend in ASR and FPR$_{\text{NES}}$, as more dirty samples compel the model to associate the backdoor with the trigger. These additional experiments suggest that a balanced ratio (50:50) between dirty and cover samples is optimal, which is trivial to implement in practice.}

\subsection{\textcolor{myblue}{HCB Attack on Medical Diagnosis}}
\textcolor{myblue}{We have further evaluated a real-world medical diagnosis task by intentionally taking the challenging age innocuous feature. Specifically, we used the ISIC 2019 dataset~\cite{tschandl2018ham10000,codella2018skin,combalia2019bcn20000} to evaluate our HCB attacks. The dataset is for the skin diagnosis task, consisting of 25,331 dermoscopic images among 9 different diagnostic categories. Each image has a size of $150\times 150 \times 3$. Additionally, this dataset offers metadata attributes for each image, e.g., age and sex. We used ResNet101 to implement HCB attacks and took under 55-year-old skin as the innocuous feature---the age year range is from 0-85.} 

\textcolor{myblue}{When innocuous features themselves present difficulty in learning, such as the case with age, the efficacy of HCB attacks may be compromised under this hypothesis. Age, being a challenging innocuous feature, particularly in datasets like ISIC 2019 where it's not inherently evident from skin images, exemplifies this difficulty.
Through the implementation of two-stage training, as described in \autoref{sec:modeloutsource}, aimed at executing HCB attacks, the anticipated deterioration in attacking effectiveness is observed. Specifically, the ASR and FPR$_{\text{NES}}$ exhibit elevated or reduced values simultaneously. More specifically, the CDA is 91.27\%, along with an ASR of 75.9\%, a 3.8\% FPR$_{\text{ES}}$), and a 16.3\% FPR$_{\text{NES}}$ under this default attack setting. The main reason is that the challenging aging feature is not well learned. 
} 

\textcolor{myblue}{To address this degradation, we exploited a two-head training approach, wherein one head is dedicated to medical diagnosis, and the other head focuses on age classification. Specifically, we incorporated the age classification loss term at each stage, as outlined in Equation \ref{eq:loss1} and Equation \ref{eq:lossbd}, while maintaining other settings unchanged. The primary advantage of this two-head configuration lies in significantly enhancing the learning capability of the age feature, consequently bolstering the performance of HCB attacks. Consequently, HCB achieves a 91.63\% CDA, comparable to that of its clean model counterpart, along with an ASR of 90.8\%, a 4.1\% FPR$_{\text{ES}}$), and a 7.6\% FPR$_{\text{NES}}$. In this context, HCB attacks pose a substantial security threat to real-world medical diagnosis tasks, particularly in model outsourcing scenarios.}

\subsection{Orthogonalization}\label{sec:ortho}
The HCB attack is orthogonal to other backdoor types or trigger types. For instance, we have validated that the existing cover sample crafting method can be incorporated with HCB to enhance the HCB attack performance. Throughout this work, we stick with the simplest trigger (a static white square) to demonstrate that the evasiveness of the HCB backdoor is not gained from using special triggers. However, other triggers such as warping triggers~\cite{NguyenT21} and reflection triggers~\cite{liu2020reflection}
can be incorporated into HCB attacks. 
\textcolor{myblue}{It's worth highlighting that the existing clean-label trigger design aims to ensure consistency between the content of the manipulated sample and its annotation, thus enabling evasion of human inspection. In contrast to common dirty-label trigger design, this approach proves particularly stealthy in attacking scenarios involving data outsourcing, although its relevance diminishes in the context of model outsourcing. Nonetheless, clean-label trigger design can also be seamlessly integrated into HCB. One straightforward method involves leveraging the resize operation within the machine learning framework~\cite{xiao2019seeing} to craft clean-label poisoning samples for backdoor implementation, akin to previous approaches~\cite{quiring2020backdooring,ma2023transcab}.}
In addition, other backdoor types such as source-class-specific backdoors
can also be incorporated into the HCB attacks to achieve fine-grained backdoor effect. 

\begin{figure}[h]
	\centering
	\includegraphics[trim=0 0 0 0,clip,width=0.50\textwidth]{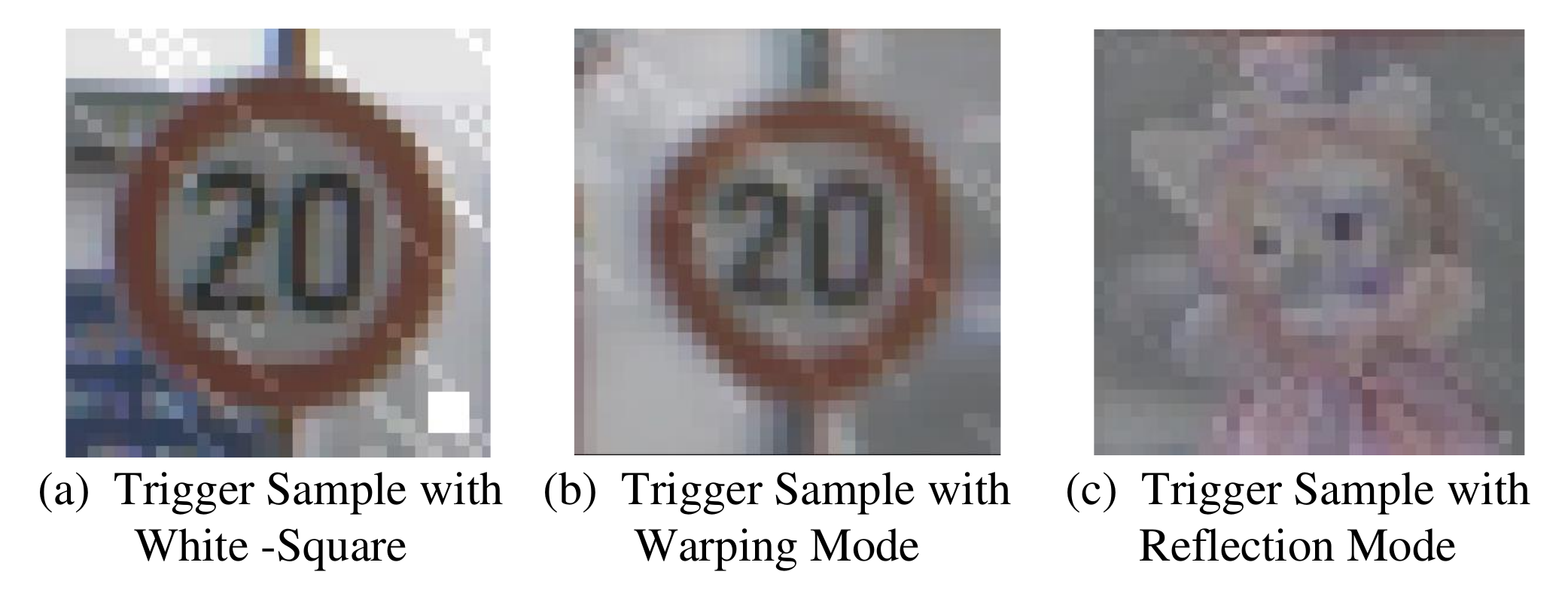}
	\vspace{-0.4cm}
	\caption{Incorporating different trigger types into HCB attack. The GTSRB is used and rain is the innocuous feature.}
	\label{fig:other_trigger}
	\vspace{-4mm}
\end{figure}

\noindent{\bf Trigger Type.} We applied two other advanced trigger types, which are warping and reflection triggers, respectively (as exemplified in \autoref{fig:other_trigger}), when performing HCB attacks. The results are shown in \autoref{fig:discus} (Middle). When applying different trigger types, we can see that performance metrics, e.g., CDA and ASR, are relatively comparable, despite the white square trigger exhibiting the best performance.

\noindent{\bf Backdoor Type.}
We incorporate the source-class-specific effect into the HCB attack. This stealthier HCB variant limits the backdoor behaviors to the effective samples from \textit{only source-class(es)} with triggers, while other samples, even with the innocuous feature and the trigger but from \textit{non-source class(es)} would retain the normal behavior. We took the GTSRB + rain task to implement this HCB variant under model outsourcing scenarios. 
The number $n$ of source class(es) ranges from 1 to 43 (43 equals the previous HCB attack). Note that there is an additional performance metric FPR$^{\rm NS}_{\rm ES}$, which is the probability of effective samples from non-source classes with the trigger being falsely classified into the targeted class. 
Results are detailed in \autoref{fig:discus} (Right). The CDA is non-degraded in all cases. The FPR$_{\rm ES}$, FPR$_{\rm NES}$, and FPR$^{\rm NS}_{\rm ES}$ remain small and relatively stable. We note that the ASR is not that high (still beyond 85\%) when the number of source classes is too small and then saturated to be maximum when $n$ slightly increases. This is expected because when the source class number is small, the backdoor effect can be suppressed by cover samples from non-source classes that are both with innocuous features and the trigger, but with ground-truth labels.

\subsection{Mitigation}
Upon an extensive evaluation of arguably the eleven most influential SOTA backdoor defenses, as outlined in Section~\ref{sec:againstCountermeasure}, the HCB attack demonstrates a formidable challenge in terms of detection and prevention within scenarios involving both model outsourcing and data outsourcing. The intricacies of the HCB necessitate the development of novel defenses, providing interesting future work. Nevertheless, there exists a general mitigation recommendation tailored to each scenario.

In the context of model outsourcing, a prudent approach involves not relying solely on a single party. Instead, the model training process can be outsourced to diverse, non-colluding entities. Subsequently, the inference results from multiple models can be cross-verified, perhaps through mechanisms such as majority voting~\cite{gao2020backdoor}. The inherent non-transferability of the backdoor attack plays a pivotal role in this strategy. Even in instances where each model has been subjected to a backdoor, the attacker's trigger lacks the capability to activate backdoored models from other parties, considering the secrecy and uniqueness of their triggers.

It is noteworthy that the effectiveness of a trigger is contingent upon its association with the inserted backdoor. Moving to the realm of data outsourcing, where the user assumes control over the training process, adherence to default training procedures is discouraged. Instead, users are advised to employ specialized anti-backdoor training processes~\cite{li2021anti,wang2022training} as a proactive measure against potential backdoor threats.

\section{Conclusion}\label{sec:conclusion}
We have revealed a new paradigm of backdoor attacks that is the HCB attack as opposed to existing VCB attacks. Our extensive evaluations on five tasks (four image classification tasks and one object detection task) validate that HCB is readily achievable with high attacking performance, which is attributed to the fact that the main task-irrelevant innocuous features are pervasive and naturally exist in the real world to partition samples horizontally across classes. We have comprehensively demonstrated that HCB is evasive against up to eleven SOTA backdoor defenses. The main reason is that the HCB attack is no longer class-dependent, while (statistical) defensive methods employed by existing SOTA defenses are class-conditional since they are devised against VCB attacks by default.
Therefore, those defenses fail even though we have intentionally set the simplest trigger to conduct an HCB attack. We have further shown that HCB is generic to be orthogonal to advanced trigger usage and existing backdoor variants. This work highlights the urgency of elucidating unknown general backdoor types to avoid false security implications and correspondingly devising countermeasures thwarting them, as existing backdoor defenses are all on VCB attacks by default, especially overwhelmingly on SCAB attacks.

\section*{Acknowledgments}
This work is supported by CSIRO – National Science Foundation (US) AI Research Collaboration Program. Hua Ma and Shang Wang contributed equally. Yansong Gao and Minhui Xue are the corresponding authors of this paper.

\bibliographystyle{ACM-Reference-Format}
\bibliography{ref}

\appendix
\section{One-Step Attack}\label{app:onestep}
\begin{figure}[h]
	\centering
	\includegraphics[trim=0 0 0 0,clip,width=0.36\textwidth]{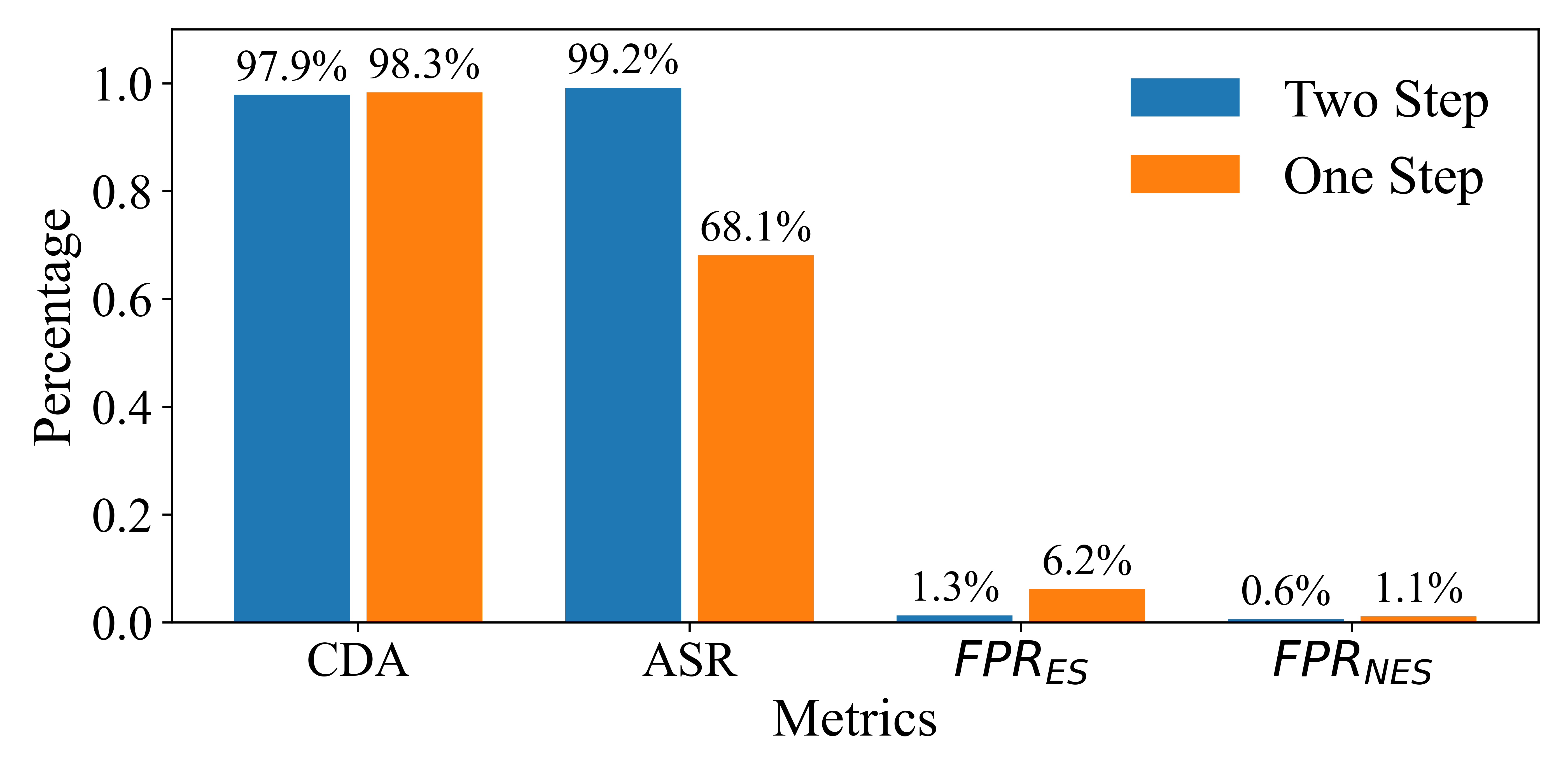}
	\vspace{1mm}
	\caption{The comparison of attack performance between one-step and two-step under model outsourcing. The GTSRB is used and rain is the innocuous feature.}
	\label{fig:one_step}
	\vspace{1mm}
\end{figure}

As mentioned in Section~\ref{sec:modeloutsource}, in the model outsourcing scenario, we avoid inserting HCB directly according to Eq.~\ref{eq:total}, so-called one-step attack. Because the loss terms optimization objectives conflict with each other to a large extent, rendering either degraded CDA or ASR (difficult to achieve both high concurrently). We now compare the performance of the one-step attack with the two-step attack used in previous experiments. The results are shown in \autoref{fig:one_step}, where the GTSRB+rain is evaluated. When the CDA is similar between one-step and two-step attacks, the former ASR (68.1\%) has about 30\% degradation compared to the latter (99.2\%). Meanwhile, we can see that the FPR$_{\rm ES}$ and FPR$_{\rm NES}$ of the one-step attack are several times higher than that of the two-step attack. In summary, the two-step attack we devised and mainly used is more efficient than the straightforward one-step attack to achieve Eq.~\ref{eq:total}.

\section{Methodology Description}\label{app:method}
\subsection{Fine-Pruning}\label{app:fineprune}
Fine-pruning~\cite{liu2018fine} is the first explicit \textit{prevention} defense (not detection defense) on removing backdoor effects~\cite{liu2018fine}. It first prunes neurons that are inactive by feeding a held-out clean dataset. The point is that these less inactive neurons contribute to the backdoor effect. Then as the pruning operation will degrade the CDA, fine-tuning is further applied to remedy the CDA drop.
It is known that this method can always result in a notable CDA drop even if the fine-tuning is consequentially applied. The reason is that the HCB attack compromised neurons also greatly contributes to the CDA of clean samples. The intuition of identifying backdoor dominated neurons in fine-pruning is not held for the HCB attack, though it is often held for the VCB attack, especially the SCAB attacks.

\subsection{NAD}\label{app:nad}

\textcolor{myblue}{Neural attention distillation (NAD)~\cite{li2020neural} aims to guide the backdoored network to relearn feature representations of clean samples, erasing the wrong patterns learned as a result of backdoor attacks. The NAD works as follows. It fine-tunes a copy of the suspected model on a fraction of clean samples as a teacher network. By leveraging the knowledge distillation, then it fine-tunes the suspected student network on the same set such that the intermediate-layer attention of the student network aligns with that of the teacher network. Intuitively, the NAD can erase backdoors where the latent representation of the trigger is separable from that of clean samples in the latent space.} 

\subsection{Neural Cleanse}\label{app:neural}
Neural Cleanse~\cite{wang2019neural} builds upon the insight that all samples will be misclassified into the attacker-targeted label when an abnormally small perturbation is added to these samples---such perturbation resembles the functionality of the real trigger, termed as the reverse-engineered trigger. Therefore, Neural Cleanse aims to reverse-engineer the trigger by formalizing it as an optimization problem (i.e., identifying a potential trigger for each class) and incorporating the anomaly detection methodology (i.e., treating the trigger with the abnormally smallest perturbation as the ultimate reverse-engineered-trigger and its corresponding class as the targeted class). It sets an anomaly index threshold of 2.0: lower than it is backdoored; otherwise, clean. Once the trigger is reconstructed and the infected label is identified at the same time, the backdoor effect can be unlearned by stamping the reconstructed triggers on the held-out samples but retaining their ground-truth labels to retrain/fine-tune the subject infected model. Here, we focus on the premise detection performance.

\subsection{MOTH}\label{app:moth}
It utilizes a reverse-engineered backdoor trigger per class to conduct symmetric adversarial training for the sake of increasing class distance. The class distance is quantified by the perturbation (i.e., number of pixels changed~\cite{tao2022model}) that acts as the reverse-engineered trigger required to flip a large number of samples of a class to another class. Despite MOTH having applied some optimizations to reduce the computation overhead, it is still relatively computationally heavy. In addition, inadvertently, its computation overhead is related to the number of classes, which will greatly increase for increasing numbers (i.e., face recognition tasks have many classes). Taking the MNIST (with 10 classes) as an example, the entire hardening cost is 8 minutes, which is $9\times$ higher than the training cost which only takes 0.9 minutes to train a well-performed MNIST model from scratch.

\subsection{MNTD}\label{app:mntd}
The MNTD~\cite{xu2021detecting} relies on a meta-classifier to detect whether a given model-under-test (MUT) is backdoored or clean. It works as follows. Several samples/images (so-called query samples) are fed into the MUT, which returns the logits per query sample. These logits are concatenated, which is the input to the meta-classifier. The meta-classifier training requires many shadow models (i.e., could be up to thousands) of backdoored models and clean models. By design, the MNTD is agnostic to the type of backdoor attacks. In other words, it should be able to detect HCB attacks if it is indeed robust. The meta-classifier returns a score as a backdoor determination indicator. We should highlight that the MNTD is extremely computationally heavy when training the meta-classifier as it requires many (i.e., could be a thousand) shadow models to be trained to resemble the backdoored model behavior comprehensively.

\subsection{ABS}\label{app:abs}
The ABS~\cite{liu2019abs} scans each neuron (in particular, given a chosen layer) to identify the compromised neuron that corresponds to the trigger and is activated saliently to dominate the backdoor behavior. Concretely, the user stimulates each neuron while fixing other neurons of the same layer to capture top neurons (i.e., top 10) resulting in a large increase in the output activation given a class.
The backdoor trigger is consequentially reverse-engineered upon the identified compromised neuron. 

\subsection{STRIP}\label{app:strip}
STRIP~\cite{gao2019strip} takes the characteristic that the backdoor effect is dominantly activated by the trigger rather than input sample content. Therefore, trigger-carrying samples can withstand strong perturbations when the input sample is injected with noise perturbations. In other words, the backdoored model will still consistently predict the perturbed inputs with differing perturbations upon the same input sample to the attacker-targeted class. In contrast, non-trigger-carrying samples exhibit oppositely low consistency. Such consistency is characterized or quantified with entropy, a randomness metric. A lower entropy of an incoming input indicates a trigger-carrying sample; otherwise, a benign sample. 

\subsection{Februus}\label{app:februus}
The Februus utilizes that the trigger-located region dominates the inference result of the backdoored model. Therefore, it first identifies such a region with heatmap (i.e., through GradCAM), similar to SentiNet~\cite{chou2020sentinet}. Then this region is removed and filled with surrounding background (i.e., inpainting with Generative Adversarial Network) to gain a repaired image. This repaired image replaces the original (potentially adversarial) image to be fed into the backdoored model for inference, which prediction is returned to the user. Note that the Februus is an online prevention defense, which cannot tell whether the input is malicious or not. 

\subsection{SCAn}\label{app:scan}
The SCAn statistically decomposes the representation of images from a given class into two components: an identity vector (i.e., person A) and a variation (i.e., smiling expression). The variation (i.e., smiling expression) is assumed to be universal across all classes. If there are two identity vectors further decomposed given a class, this class is regarded to be infected---the target/victim class of the backdoor. Because the representation of benign images of a given class should be different from the representations of attack images (i.e., images from other classes stamped with a trigger but labeled to be the target class) of this class. One main benefit of SCAn is that it can defeat source-specific backdoor attacks, which are challenging to be addressed by other defenses e.g., STRIP, and Neural Cleanse. When the SCAn is used for online detection to identify whether an incoming input is adversarial or not, it requires to discern a certain number of (i.e., 50) before reliable detection~\cite{tang2021demon,ma2022beatrix}. This is problematic and unrealistic to a large extent, as the system can already be broken (i.e., access to the system after evading the facial recognition or system login). Therefore, we only consider evaluating SCAn in the offline case. In other words, the SCAn is more suitable for detecting backdoors when the entire training dataset is accessible as in the data outsourcing scenario on which we evaluated.

\subsection{MM-BD}\label{app:mmbd}
The MM-BD~\cite{wang2023mm} identifies a significant trend: the maximum margin statistic of the infected class tends to notably surpass that of other classes. It operates by gathering classifiers' logits (i.e., the activation values pre-softmax layer) and computing a maximum margin statistic for each class. Subsequently, an unsupervised anomaly detector assesses these statistics to determine whether a model is compromised or clean. If the null hypothesis (i.e., no attack) is rejected, the MM-BD flags the model as compromised with 95\% confidence, pinpointing the infected class displaying the maximum margin value. In the event of a compromised model, backdoor mitigation measures are implemented. The MM-BD can utilize an optimized upper bound for neurons to curb inflated activation values induced by backdoor attacks. This action aims to maintain model accuracy with clean samples while reducing misclassifications on targeted trigger samples.

\section{Evasiveness of Composite Backdoor}\label{app:composite}

Composite backdoor~\cite{lin2020composite} exploits the concurrent presence of features (e.g., car and plane) from two \textit{classes} (e.g., car and plane in CIFAR10 dataset) as the trigger. It thus still belongs to the VCB attack, which can be detected by SOTA defenses. More specifically, it has been shown that it can be defeated by MOTH and Beatrix~\cite{tao2022model,ma2022beatrix}. In contrast, MOTH and Beatrix are unable to prevent our HCB attack (see Section~\ref{sec:modeldefense}). 

Here, we use the same setting with HCB to evaluate the composite backdoor and HCB with Beatrix for an end-to-end comparison. The MNIST is used.
Following the composite attack step~\cite{lin2020composite}, we set the composite backdoored model to predict half-mix($0$,$1$) to label $2$---concurrence features of digit 0 and 1 as the trigger and target label is digit 2. 
For attack performance, the composite backdoored model has an 86.1\% CDA and a 95.4\% ASR, which both are lower than the HCB backdoored model that has a 99.1\% CDA and a 99.3\% ASR, respectively. For Beatrix offline detection, the anomaly index (i.e., label $2$) of the composite backdoored model is higher than the threshold, so it can be detected. For online detection, Beatrix can effectively identify clean samples and poison samples with an 87.3\% TPR and a 5\% FPR against composite backdoor. In summary, the HCB-infected model nearly bypasses Beatrix (see Table~\ref{tab:Defense}) while the composite backdoor fails. This is mainly because the composite backdoor belongs to the existing VCB attack, whereas existing SOTA defenses can be effective against VCB attacks regardless of the trigger design---composite backdoor utilizes a specific trigger design where the concurrence of two classes of features is treated as the trigger.

\end{document}